\documentclass[pre,preprint,superscriptaddress,amsmath,amssymb]{revtex4-1}
\usepackage{graphicx,subfigure,color}
\usepackage{dcolumn,amssymb,amsthm}
\usepackage{bm,amsmath,longtable,booktabs, array,rotfloat}
\usepackage{nomencl,multirow,longtable}
\newtheorem{remark}{Remark}

\begin{document}


\title{Curved boundary conditions of the lattice Boltzmann method for simulating microgaseous flows in the slip regime}

\author{Liang Wang}
\affiliation{Beijing Key Laboratory of Emission Surveillance and Control for Thermal Power Generation, North China Electric Power University, Beijing 102206, China}
\affiliation{School of Energy Power and Mechanical Engineering, North China Electric Power University,
Beijing 102206, China}
\author{Shi Tao}
\affiliation{Guangdong Provincial Key Laboratory of Distributed Energy Systems, Dongguan University of Technology, Dongguan 523808, China}
\author{Junjie Hu}
\affiliation{Faculty of Engineering, China University of Geosciences, Wuhan 430074, China}
\author{Kai Zhang}
\affiliation{Beijing Key Laboratory of Emission Surveillance and Control for Thermal Power Generation, North China Electric Power University, Beijing 102206, China}
\affiliation{School of Energy Power and Mechanical Engineering, North China Electric Power University,
Beijing 102206, China}
\author{Gui Lu}
 \email[Corresponding author:\quad]{lugui02@gmail.com}
 \affiliation{Key Laboratory of Power Station Energy Transfer Conversion and System of Ministry of Education, North China Electric Power University, Beijing 102206, China}

\date{\today}

\begin{abstract}
The lattice Boltzmann method (LBM) has shown its promising capability in simulating microscale gas flows. However, the suitable boundary condition is still one of the critical issues for the LBM to model microgaseous flows involving curved geometries. In this paper, a local boundary condition of the LBM is proposed to treat curved solid walls of microgaseous flows. The developed boundary treatment combines the Maxwellian diffuse reflection scheme and a single-node boundary scheme which contains a free parameter as well as the distance ratio. The curved boundary condition is analyzed within the multiple-relaxation-time (MRT) model for a unidirectional microflow. It is shown that the derived slip velocity depends on the free parameter as well as the distance ratio and relaxation times. By virtue of the free parameter, the combination parameter and the uniform relaxation time are theoretically determined to realize the accurate slip boundary condition. In addition, it is found that besides the halfway diffuse-bounce-back (DBB) scheme, previous curved boundary schemes only containing the distance ratio cannot ensure uniform relaxation times to realize the slip boundary condition. Some numerical examples with planar and curved boundaries are carried out to validate the present curved boundary scheme. The good and robust consistency of numerical predictions with analytical solutions demonstrates our theoretical analysis.
\end{abstract}

\maketitle

\section{\label{intro}Introduction}
With the rapid progress of microscience and nanotechnology, microscale gaseous flows have attracted increasing research attention in recent years \cite{Ho98,Karniadakis02,Zhang12}. For such gas flows, the mean free path of gas moleculars ($\lambda$) is usually comparable to the characteristic length scale of the flow system ($H$). As such, the gas flows are far from the thermodynamical equilibrium, and the classical Navier-Stokes equations based on the continuum assumption are no longer valid. Generally, the Knudsen number $\text{Kn}=\lambda/H$ is used to measure the departure degree of microscale gaseous flows from the classic hydrodynamic theory.
Based on the magnitude of $\text{Kn}$, gas flows could be classified into four major regimes: the continuum flow regime with $\text{Kn}\leq0.001$, the slip flow regime with $0.001<\text{Kn}\leq0.1$, the transition flow regime with $0.1<\text{Kn}\leq10$, and the free-molecular flow regime with $\text{Kn}>10$. It is widely accepted that the Boltzmann equation can work for rarefied gas flows with arbitrary Knudsen numbers \cite{Cercignani90}. Therefore, as a discrete scheme derived from the Boltzmann equation \cite{HeX98,ShanX98}, the lattice Boltzmann method (LBM) has been believed to have the potential for simulating microscale gaseous flows. Since being applied to microscale gaseous flows first in 2002 \cite{Nie02,Lim02}, the LBM has received increasing interests over the past dozen years \cite{Shu05,ZhangY05,ZhangJ05,Kima08}.

With the increase of $\text{Kn}$, the Navier-Stokes equations with the no-slip boundary condition become invalid for gas flows, while the fluid slippage on a solid-fluid boundary will arise due to the gaseous nonequilibrium effect \cite{Sone02,Sharipov11}. Therefore, for the LBM to capture the gaseous slip phenomenon in simulating microgaseous flows, the boundary condition is of critical importance for effectively predicting gas-solid interactions. This important issue has attracted substantial researches toward accurate slip boundary treatments. Nie \emph{et al}. \cite{Nie02} employed the standard bounce-back (BB) boundary condition to predict the flows in microchannels, and found a nonzero velocity proportional to the square of $\text{Kn}$ on the channel wall. While, the slip velocity was later revealed to be a numerical artifact \cite{Verhaeghe09} actually. Lim \emph{et al.} \cite{Lim02} investigated the specular reflection (SR) boundary condition in microchannel flow simulations, and the slip velocity was not well consistent with some existing analytical solutions. Ansumali and Karlin \cite{Ansumali02} applied the Maxwellian diffusive (MD) boundary condition for the Kramer's problem, but the scheme tends to overpredict the slip velocity. Therefore, as revealed in the subsequent researches, the above pure boundary schemes cannot accurately capture the slip phenomenon.

The shortcomings in the above scenario have thus stimulated improving hybrid schemes which combine the pure boundary conditions with an accommodation coefficient \cite{Verhaeghe09,Succi02,Sbragaglia05,Tang05}. By mixing the BB and the SR boundary conditions, Succi \cite{Succi02} proposed a hybrid scheme, i.e., the bounce-back and specular reflection (BBSR) scheme for simulating microscale flows. A generalization of the BBSR scheme was subsequently developed and analyzed by Sbragaglia and Succi \cite{Sbragaglia05}. Tang \emph{et al.} \cite{Tang05} proposed the diffusive and the specular reflection (DSR) scheme, which comes from the combination of the MD and SR boundary schemes. Thanks to the tunable accommodation parameter, the degree of slip can be freely controlled to recover different slip models in the two hybrid schemes. Another hybrid boundary scheme is the diffusive and bounce-back (DBB) scheme \cite{Verhaeghe09,Chai08}, which is a combination of the MD and the BB boundary conditions. Owing to the advantage of local computation superior to the BBSR and DSR schemes, the DBB scheme may possess more potential in simulating microscale gaseous flows with complex geometries. For hybrid slip boundary conditions, how to choose the combination parameter is not a convenient task. With the Bhatnagar-Gross-Krook (BGK) model and the multiple-relaxation-time (MRT) model, Guo \emph{et al.} \cite{Guo07,Guo08} successively analyzed the discrete effects of the BBSR and DSR boundary conditions, and found the interrelationship between the two schemes. To realize the accurate slip boundary condition, they proposed a strategy to determine the combination parameter. Verhaeghe \emph{et al.} \cite{Verhaeghe09} and Chai \emph{et al.} \cite{Chai08,Chai10} further mathematically analyzed the discrete effect of the DBB scheme. Their results illustrated that the combination parameter should be chosen carefully to impose the accurate slip boundary condition.
Noteworthily, these hybrid boundary schemes are originally designed to implement the wall location with a definite distance between lattice nodes. Szalm$\acute{a}$s \cite{Szal06} used an interpolated method to combine the BB and the SR schemes, and it permits arbitrary locations of the slip wall. However, this slip boundary condition is proposed for microgaseous flows specially with straight walls. Therefore, for microscale slip flows with curved boundaries, which bring variable wall locations in relation to the underlying grid, the above-mentioned studies theoretically cannot yield enough accurate results especially at a small grid resolution.

For a more accurate treatment of curved boundaries in microgaseous flows, there have been developed a number of boundary conditions considering the actual boundary shape in the literature. Suga \cite{Suga13} introduced an interpolation method into the DBB boundary condition for microscale flow simulations. The effect of curved surface is represented by its intersection distance ratio between neighbouring lattice nodes. But, the combination coefficient is still determined by that from the halfway bounce-back case. Based on the non-equilibrium extrapolation method for curved boundaries and the counter-extrapolation method for the velocity/temperature at curved surfaces, Liu \emph{et al.} \cite{LiuZ19} proposed a boundary condition involving the distance ratio for thermal gaseous microflows with curved slip walls. However, as revealed in the literature \cite{Tao15,Silva17,Silva18}, the discrete effects also exist in curved boundary treatments for slip walls, which must be minimized to capture correct microgaseous slip phenomenon. Within the MRT model, Tao and Guo \cite{Tao15} incorporated the effect of distance ratio to analyze the DBB scheme, and then developed a boundary scheme to realize the slip boundary condition at curved boundaries. Silva and Semiao \cite{Silva17,Silva18} introduced the multireflection framework to gaseous slip flows, and put forward curved slip boundary schemes within the two-relaxation-time (TRT) model. The theoretical analysis on curved boundary schemes revealed that the numerical accuracy is related with the combination parameter and the relaxation times, which are functions of the wall cut-link distance ratio \cite{Tao15,Silva17,Silva18}. Even for microscale binary gaseous flows, such results are also exposed as the DBB scheme is analyzed to derive the slip velocity at curved surfaces \cite{Ren19}. In the LBM for microflows with curved walls, previous studies have indicated that the combination parameter can change locally with the wall cut-link distance ratio. However, the relaxation parameter should also vary with the distance ratio to realize the slip boundary condition, which drives the anisotropic collision operator. Unfortunately, such issue cannot be overcome with previous curved boundary schemes which only contain the distance ratio.

To attain uniform relaxation parameters as noted above, one natural strategy is to impose the halfway boundary scheme for curved slip walls, and the actual locations between lattice nodes are approximated as halfway wall locations \cite{Chai10}. In this way, the uniform relaxation parameter can be then determined by the distance ratio fixed at $0.5$. Following this line of thought, there have been some woks reported to treat the curved slip boundary by the halfway DBB scheme \cite{Verhaeghe09,Chai10,Guo11}.
However, the real curved geometry with the halfway approximation will lost its fidelity under coarse grid resolutions, and undesired errors may arise to contaminate the simulation accuracy \cite{Silva17,Silva18}. Therefore, based on the above literature review, it promotes us to resolve such a critical issue for microgaseous flows with curved boundaries: how to retain the relaxation parameters unchanged to realize the slip boundary condition. Accordingly, the present work has the following twofold objectives: to develop a curved kinetic boundary condition which involves additional parameters besides the distance ratio; and to realize the accurate slip boundary condition while with invariant relaxation parameters.

In this work, a kinetic boundary condition is first constructed by combining an interpolation-based scheme and the diffuse reflection rule. Different from previous curved boundary conditions, the present scheme is inspired by the idea in Ref. \cite{Zhao19} to include a free parameter besides the distance ratio, which can bring infinitely many curved boundary schemes for microgaseous flows. As the subsequent step, the boundary scheme is mathematically analyzed within the MRT model for a unidirectional flow. On the basis of theoretical derivations, the combination parameter and a strategy to ensure uniform relaxation parameters are given to realize the slip boundary condition at curved slip walls. Numerical simulations are then carried out to validate the developed method in the cases of aligned and inclined flat walls and curved walls. As compared with the halfway DBB boundary scheme and other curved boundary schemes with several specific free parameters, the present method can predict the most accurate results consistent with the analytical solutions.

\section{MRT LBE for microscale gaseous flows }\label{Sec2}
The LBM is derived from the discretization of continuous Boltzmann equation in both time, space and velocity space. The discrete velocity distribution functions evolve according to the following lattice Boltzmann equation (LBE),
\begin{equation}\label{LBENS}
  f_i(\bm{x}+\bm{c}_i\delta_t, t+\delta_t)-f_i(\bm{x}, t)=\Omega_i(f)(\bm{x}, t)+\delta_t F_i(\bm{x}, t), \quad i=0,1,\cdots, b-1,
\end{equation}
where $ f_i(\bm{x}, t)$ is the distribution function associated with the discrete velocity $\bm{c}_i$ at position $\bm{x}$ and time $t$, $\delta_t$ is the time step, $\Omega_i(f)$ is the discrete collision operator, and $F_i$ is the discrete forcing term, and $b$ is the number of discrete velocities.

The BGK or single-relaxation-time model is the most widely used collision operator in the LBM. However, the slip velocity derived within the BGK model depends on the relaxation time, which is grid resolution dependent for a given $\text{Kn}$ \cite{Verhaeghe09,Guo08}. This means that some unphysical numerical artifacts besides the physical part exist in the slip velocity. To avoid this problem, in this work we turn to employ the MRT collision operator adhered to the LBE, which is written as
\begin{equation}\label{LBEColli}
   \Omega_i(f)=-\sum_j(\bm{M}^{-1}\bm{S}\bm{M})_{ij}\left[f_j-f_j^{(\text{eq})}\right],
\end{equation}
where $\bm{M}$ is a $b\times b$ transformation matrix, which maps $f_i$ onto the moment space via $\bm{m}=\bm{M}\bm{f}$ with $\bm{f}=(f_0,f_1,\cdots,f_{b-1})^\text{T}$, $\bm{S}=\text{diag}(\tau_0, \tau_1,\cdots, \tau_{b-1})^{-1}$ is a diagonal relaxation matrix with its non-negative element $\tau_i$ being the relaxation time for the $i$-th moment. $f_j^{(\text{eq})}$ is the equilibrium distribution function which is dependent on the gas density $\rho$, velocity $\bm{u}$ and temperature $T$,
\begin{equation}\label{LBEequi}
 f_j^{(\text{eq})}=\omega_j\rho\left[1+\frac{\bm{c}_j\cdot\bm{u}}{c_s^2}+\frac{(\bm{c}_j\cdot\bm{u})^2}{2c_s^4}
    -\frac{\bm{u}^2}{2c_s^2}\right], \quad j=0,1,\cdots, b-1,
\end{equation}
where $\omega_j$ is the weight coefficient, $c_s=\sqrt{RT}$ ($R$ is the gas constant) is the lattice sound speed. For isothermal flows, $c_s$ is determined by the lattice speed $c=\delta_x/\delta_t$ with $\delta_x$ being the lattice spacing. For the discrete forcing term $F_i$ in Eq. \eqref{LBENS}, it should be taken as \cite{Guob13}
\begin{equation}
  \bm{F}=\bm{M}^{-1}\left(\bm{I}-\frac{\bm{S}}{2}\right)\bm{M}\bm{\overline{F}},
\end{equation}
where $\bm{I}$ is the identity matrix, $\bm{F}=(F_0,F_1,\cdots,F_{b-1})^\text{T}$, and $\bm{\overline{F}}=(\overline{F}_0,\overline{F}_1,\cdots,\overline{F}_{b-1})^\text{T}$ is expressed as
\begin{equation}\label{LBEforce}
   \overline{F}_i= \omega_i\rho\left[\frac{\bm{c}_j\cdot\bm{a}}{c_s^2}+\frac{\bm{u}\bm{a}:(\bm{c}_i\bm{c}_i-c_s^2\bm{I})}{c_s^4}\right],
\end{equation}
where $\bm{G}=\rho\bm{a}$ is the external force.

In this work, we employed the two-dimensional nine-velocity (D2Q9) model, where the discrete velocities $\bm{c}_i$ are defined by
\begin{equation}\label{Eq3}
 \bm{c}_i:=c\bm{e}_i=
  \begin{cases}
  c(0,0),&i=0,\\
  c\bigl(\cos\bigl[(i-1)\pi/2\bigr],\sin\bigl[(i-1)\pi/2\bigr]\bigr),&i=1,2,3,4,\\
  \sqrt{2}c\bigl(\cos\bigl[(i-1)\pi/2+\pi/4\bigr],\sin\bigl[(i-1)\pi/2+\pi/4\bigr]\bigr),&i=5,6,7,8,
  \end{cases}
\end{equation}
where $c=\delta_x/\delta_t$ with $\delta_x$ denoting the lattice spacing. Correspondingly, the sound speed $c_s=\sqrt{RT}=c/\sqrt{3}$, and the weight coefficients are given by $\omega_0=4/9$, $\omega_{1-4}=1/9$ and $\omega_{5-8}=1/36$. Via the Gram-Schmidt orthogonalization procedure on the discrete velocities $\bm{c}_i$, there are different versions of transformation matrices $\bm{M}$, and one form of $\bm{M}$ as $c=1$ is given by \cite{Lallemand00}
\begin{align}
\bm{M}=
  \left(
  \begin{array}{ccccccccc}
    1  & 1 & 1 & 1 & 1 & 1 & 1 & 1 & 1\\
   -4 & -1 & -1 & -1 & -1 & 2 & 2 & 2 & 2\\
   4 & -2 & -2 & -2 & -2 & 1 & 1 & 1 & 1\\
   0 &  1 & 0  & -1 & 0  & 1 & -1 & -1 & 1\\
   0 & -2 & 0 & 2 & 0 & 1 & -1 & -1 & 1\\
   0 & 0 & 1 & 0 & -1 & 1 & 1 & -1 & -1\\
   0 & 0 & -2 & 0 & 2 & 1 & 1 & -1 & -1\\
   0 & 1 & -1 & 1 & -1 & 0 & 0 & 0 & 0\\
   0 & 0 & 0 & 0 & 0 & 1 & -1 & 1 & -1
  \end{array}
  \right).
\end{align}
As a result, the nine discrete velocity moments $m_i$ from the distribution functions $f_i$ are expressed as
\begin{equation}
   \bm{m}=\bm{M}\bm{f}=(\rho, e, \varepsilon, j_x, q_x, j_y, q_y, p_{xx}, p_{xy})^\text{T}.
\end{equation}
The corresponding relaxation matrix $\bm{S}$ for the nine moments is written as
\begin{equation}
   \bm{S}=\text{diag}(\tau_\rho, \tau_e, \tau_\varepsilon, \tau_j, \tau_q, \tau_j, \tau_q, \tau_s, \tau_s)^{-1}.
\end{equation}

The fluid density $\rho$ and velocity $\bm{u}=(u,~v)$ are respectively defined as the zeroth and first-order moments of $f_i$
\begin{equation}\label{FluidMacr}
  \rho=\sum_i f_i, \quad \rho\bm{u}=\sum_i \bm{c}_i f_i+\frac{\delta_t}{2}\rho\bm{a}.
\end{equation}
Through the Chapman-Enskog or linear analysis, the Navier-Stokes equations can be derived from the MRT-LBE model \eqref{LBENS}. The fluid pressure $p$ is determined by $p=c_s^2\rho=\rho RT$, and the shear and bulk viscosities are respectively given by
\begin{equation}\label{EqViscos}
  \nu = c_s^2\left(\tau_s-\frac{1}{2}\right)\delta_t, \qquad \zeta=c_s^2\left(\tau_e-\frac{1}{2}\right)\delta_t.
\end{equation}

The numerical implementation of Eq. \eqref{LBENS} is divided into two steps:
\begin{align}
   &\text{Collision}:\quad f_i'(\bm{x},t)=f_i(\bm{x},t)-\sum_j(\bm{M}^{-1}\bm{S}\bm{M})_{ij}\left[f_j(\bm{x}, t)-f_j^{(\text{eq})}(\bm{x}, t)\right]+\delta_t F_i(\bm{x}, t),\notag\\
   &\text{Streaming}:\quad f_i(\bm{x}+\bm{c}_i\delta_t, t+\delta_t)=f_i'(\bm{x},t),
  \end{align}
where $f_i'(\bm{x},t)$ is the post-collision distribution function. In the framework of MRT model, the collision step is usually implemented in the moment space
\begin{equation}\label{MomentPost}
  \begin{split}
   \bm{m}'(\bm{x},t)&=\bm{m}(\bm{x},t)-\bm{S}\left[\bm{m}(\bm{x},t)-\bm{m}^{(\text{eq})}(\bm{x},t)\right]+\delta_t \bm{\hat{F}}(\bm{x}, t),\\
   \bm{f}'(\bm{x},t)&=\bm{M}^{-1}\bm{m}'(\bm{x},t),
  \end{split}
\end{equation}
where $\bm{m}':=\bm{M}\bm{f}'$ with $\bm{f}'=(f_0',f_1',\cdots,f_8')^\text{T}$ is the post-collision moment, $\bm{m}^{(\text{eq})}:=\bm{M}\bm{f}^{(\text{eq})}$ and $\bm{\hat{F}}:=\bm{M}\bm{F}=\left(\bm{I}-\frac{\bm{S}}{2}\right)\bm{M}\bm{\overline{F}}$ denote the equilibria and forcing term in the moment space, respectively. After the collision step is completed, the streaming step is then executed by transforming the moments back to the velocity space.

In contrast to continuum flows, the LBM for microscale flows should consider the rarefaction effects, which are commonly characterized by the dimensionless Knudsen number. The most important characteristic parameter in continuum flows is the Reynolds number $\text{Re}$, which determines the relaxation time $\tau_s$ in the simulations. However, for microscale gas flows, the relationship between $\tau_s$ and $\text{Kn}$ should be carefully established to ensure the consistency criterion \cite{Guo06}, which is read as
\begin{equation}\label{tKnrela}
   \tau_s=\frac{1}{2}+\sqrt{\frac{6}{\pi}}\text{Kn}\frac{H}{\delta_x}.
\end{equation}

Another important issue for the LBM applied to microscale flows is the kinetic boundary condition. As reviewed previously, there have been some boundary conditions proposed for the LBE to simulate microgaseous flows, such as the BBSR scheme, DSR scheme and DBB scheme. These kinetic boundary conditions are originally designed for the case of flat walls with definite locations between lattice nodes. Furthermore, it has been exposed in the literature \cite{Guo07,Guo08} that some discrete effects exist in the boundary schemes and should be corrected to realize the slip boundary condition. For a more accurate treatment of curved walls, it is \emph{common} to include the distance ratio in the boundary condition to preserve the actual shape of curved geometries \cite{Suga13,LiuZ19}. However, when these curved boundary schemes are implemented to realize the slip boundary condition at curved walls, the relaxation time in the LBE will depend on the distance ratio and thus varies with link directions at different boundary nodes \cite{Tao15,Silva17,Silva18}. Noteworthily, such a problem cannot be overcome in previous curved boundary schemes only containing the distance ratio.

\section{Kinetic boundary condition for curved slip walls}\label{Sec3}
To remedy the above issue of nonuniform relaxation parameters, one natural and promising way is to introduce additional parameters besides the distance ratio into a kinetic boundary condition. As a matter of fact, by resorting to a single-node boundary scheme \cite{Zhao19} for a free parameter besides the distance ratio, we recently overcome the numerical slip at curved no-slip walls \cite{WangL20} with invariant relaxation parameters. Thus, it motivates and inspires us to construct such a parameterized boundary condition for microflows with curved walls, and then generalize the strategy to ensure uniform relaxation parameters from continuum flows to microgaseous flows.

\subsection{Curved boundary scheme for microscale gaseous flows}
\begin{figure}
\centering
\includegraphics[width=0.78\textwidth,height=0.26\textheight]{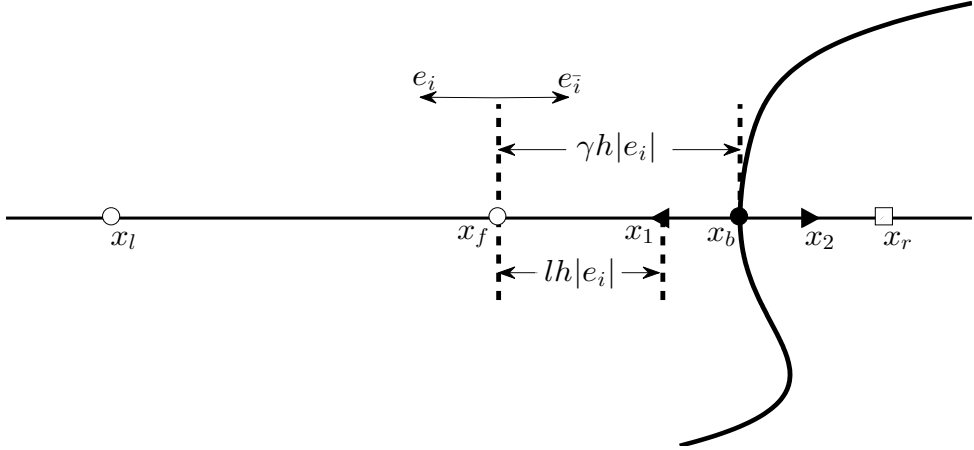}
\caption{Schematic of a curved-wall boundary along one single lattice direction. The thin solid line is the grid line, and the thick curved one represents the boundary surface. White circles ($\circ$): the fluid nodes; Black circle ($\bullet$): the intersection point of the boundary with the grid line; Square box ($\square$): the solid node outside the computational domain.}
\label{Fig:ScheSlipB}
\end{figure}

We now propose a kinetic boundary condition that involves an additional parameter besides the distance ratio as noted above. Because a curved boundary in simulations can be modeled separately for each lattice direction, we consider a single direction $\bm{e}_i$ pointing from the boundary surface at $\bm{x}_b$ to the boundary node $\bm{x}_f$, as shown in Fig. \ref{Fig:ScheSlipB}.
The surface point $\bm{x}_b$ is intersected by $\bm{x}_f$ and the solid node $\bm{x}_r$, and then the actual location of curved boundary can be depicted by the distance ratio $\gamma$ as $\gamma=|\bm{x}_b-\bm{x}_f|/|\bm{x}_r-\bm{x}_f|$. At the boundary node $\bm{x}_f$, the unknown distribution function $f_i(\bm{x}_f,t+\delta_t)$ in the LBM is specified according to the kinetic boundary condition, which aims to realize the slip velocity condition at physical walls. Noteworthily, for microscale flows with curved boundaries or in complex geological porous media, a local boundary condition that involves the current information at boundary nodes is desired especially. In the present work, our construction aims to a local curved boundary condition for microscale gas flows.

Recall that the DBB boundary condition, which has been developed for both planar and curved slip walls \cite{Verhaeghe09,Chai08,Tao15}, pertains to a local boundary scheme. This is due to its combination of the BB and the MD schemes, which is formulated for the unknown distribution function as follows:
\begin{equation}\label{EqDBB}
 f_i(\bm{x}_f,t+\delta_t)=r\left[f_{\bar{i}}'(\bm{x}_f, t)+2\omega_i\rho_f\frac{\bm{c}_i\cdot\bm{u}_b}{c_s^2}\right]+(1-r)f_i^{(\text{eq})}(\rho_f,\bm{u}_b),
\end{equation}
where the subscript $\bar{i}$ indicates $\bm{e}_{\bar{i}}=-\bm{e}_i$, $\rho_f=\rho(\bm{x}_f, t)$ and $\bm{u}_b=\bm{u}(\bm{x}_b,t)$ is the wall velocity. The combination parameter $r$ represents the bounce-back fraction and ranges in the region of $0\leq r \leq 1$. Only the lattice directions of $\bm{e}_i$ and $\bm{e}_{\bar{i}}$ are involved at the current node $\bm{x}_f$, and thus the DBB scheme shares the perfect feature of local computation for curved walls. For this local scheme, the bounce-back part accounts for the no-slip boundary condition, while the Maxwell diffuse part is responsible for the gas slippage at solid walls. On the other hand, Zhao \emph{et al}. \cite{Zhao19} recently proposed a single-node boundary scheme, which contains a free parameter besides $\gamma$, for the no-slip boundary condition. Later, its ability to overcome the discrete effect with uniform parameters has been revealed in our recent work \cite{WangL20}. Along with the underlying structure of the DBB scheme, we thus replace the bounce-back part in Eq. \eqref{EqDBB} by the scheme in Ref. \cite{Zhao19}, and derive the following boundary condition
\begin{align}\label{EqNewBDSC}
 f_i(\bm{x}_f,t+\delta_t)&=r\left[\frac{1+l-2\gamma}{1+l}f_{\bar{i}}(\bm{x}_f, t)+\frac{l}{1+l}f_i'(\bm{x}_f, t)+\frac{2\gamma-l}{1+l}f_{\bar{i}}'(\bm{x}_f,t)+\frac{2}{1+l}\omega_i\rho_f\frac{\bm{c}_i\cdot\bm{u}_b}{c_s^2}\right] \notag\\
 &\qquad +(1-r)f_i^{(\text{eq})}(\rho_f,\bm{u}_b),
\end{align}
where $l$ is a free parameter besides $\gamma$ such that $\bm{x}_b=(\bm{x}_1+\bm{x}_2)/2$ (see Fig. \ref{Fig:ScheSlipB}). Clearly, the boundary condition \eqref{EqNewBDSC} is a local boundary scheme, and can preserve the geometry fidelity of curved walls by the distance ratio $\gamma$. Since the no-slip boundary scheme in Ref. \cite{Zhao19} is incorporated, the present boundary condition is inherently associated with the diffusive scaling $\delta_t=\eta h^2,~h=\delta_x$ ($\eta$ is an adjustable parameter). Because $l$ takes values along the same lattice direction as $\gamma$ does (cf Fig. \ref{Fig:ScheSlipB}), it can be considered as a function of $\gamma$. In addition, the parameter $l$ ranges in $\text{max}\{0, 2\gamma-1\}\leq l\leq 2\gamma$ to ensure the convex combination of distribution functions for the no-slip part.

In the LBM for microscale flows, previous curved boundary schemes only contain the distance ratio $\gamma$, while the present boundary scheme  \eqref{EqNewBDSC} introduces an adjustable parameter $l$ besides $\gamma$. With the variable parameter $l$, the boundary scheme \eqref{EqNewBDSC} can hence bring numerous boundary conditions for microgaseous flows. More importantly, we will show later that owing to the free parameter $l$, uniform relaxation parameters can be fulfilled to realize an exact prescribed slip boundary condition. Some remarks about the boundary curved scheme are given as follows:
\begin{remark}\label{r1Rem}
When $r=1$, the boundary scheme \eqref{EqNewBDSC} will degenerate to the single-node scheme proposed in Ref. \cite{Zhao19} for continuum flows with no-slip walls. From this viewpoint, the present scheme is the generalized version of that in Ref. \cite{Zhao19}.
\end{remark}
\begin{remark}
When $l=0$ and $\gamma=\frac{1}{2}$, the boundary scheme \eqref{EqNewBDSC} then degenerates to the DBB scheme \eqref{EqDBB}. This also confirms the intrinsic halfway consumption of the bounce-back part adhered to the boundary scheme \eqref{EqDBB}.
\end{remark}
\begin{remark}
In the present work, the effect of curved wall geometry is directly embodied in the boundary condition \eqref{EqNewBDSC} by the distance ratio $\gamma$. While in Ref. \cite{Tao15}, it is absorbed in the DBB scheme \eqref{EqDBB} through the combination parameter to realize the slip boundary condition at curved walls.
\end{remark}

As noted before, the combination parameter $r$ in hybrid boundary schemes plays the vital role and directly affects the degree of slip at physical boundaries \cite{Guo07,Guo08}. Therefore, it should be carefully chosen to derive reasonable simulation results. Previous studies with the MRT model have shown that $r$ is related with sever factors, including the relaxation time, the gas-solid interaction parameter and the distance ratio \cite{Guo08,Tao15}. For the present boundary scheme \eqref{EqNewBDSC} with the free parameter $l$, the influence of $r$ is expected to be more complicated in treating curved slip walls. Next, we will conduct a theoretical analysis to investigate how to specify $r$ here, and simultaneously ensure the above-mentioned uniform relaxation parameters.

\subsection{Analysis of the curved boundary scheme}
To simplify the analysis, we consider the steady incompressible Poiseuille flow in a microchannel, which has been extensively employed for theoretical analysis in previous studies \cite{Shu05,Verhaeghe09,Chai08,Guo08,Guo11,Tao15}.
\begin{figure}
\centering
\includegraphics[width=0.78\textwidth,height=0.3\textheight]{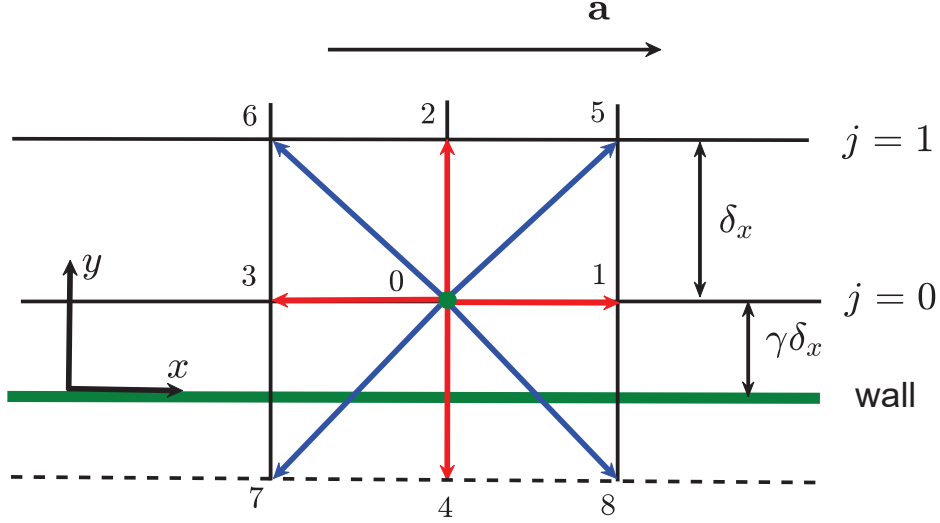}
\caption{Schematic of the flow and lattice arrangement with a distance ratio $\gamma$. The wall boundary corresponding to the halfway DBB boundary condition is placed with $\gamma=1/2$.}
\label{ScheMiclattice}
\end{figure}
As sketched in Fig. \ref{ScheMiclattice}, the flow is driven by a constant force $\rho\bm{a}=\rho(a,~0)$ along the $x$ direction, and is subjected to the following assumptions:
\begin{equation}\label{EqAssumpt}
   \rho=\text{const}, \quad v=0, \quad \partial_x\phi=0, \quad \partial_t\phi=0,
\end{equation}
where $\phi$ is an arbitrary flow variable. The lattice node at the layer $j=0$ is located with an arbitrary distance $\gamma\delta_x$ away from the solid wall. Here $j$ is the index of the grid line at $y_j=(j+\gamma)\delta_x$. After the streaming step, the unknown distribution functions, $f_2$, $f_5$ and $f_6$ at $j=0$ are determined according to the boundary scheme \eqref{EqNewBDSC}:
\begin{subequations}\label{EqMicroBound}
 \begin{equation}\label{fb2}
    f_2^0=r\left[\frac{1+l-2\gamma}{1+l}f_4^0+\frac{l}{1+l}f_2^{'0}+\frac{2\gamma-l}{1+l}f_4^{'0}
    +\frac{2}{1+l}\omega_2\rho\frac{\bm{c}_2\cdot\bm{u}_b}{c_s^2}\right]+(1-r)f_2^{(\text{eq})}(\rho,u_w),
  \end{equation}
  \begin{equation} \label{fb5}
    f_5^0=r\left[\frac{1+l-2\gamma}{1+l}f_7^0+\frac{l}{1+l}f_5^{'0}+\frac{2\gamma-l}{1+l}f_7^{'0}
    +\frac{2}{1+l}\omega_5\rho\frac{\bm{c}_5\cdot\bm{u}_b}{c_s^2}\right]+(1-r)f_5^{(\text{eq})}(\rho,u_w),
  \end{equation}
  \begin{equation} \label{fb6}
    f_6^0=r\left[\frac{1+l-2\gamma}{1+l}f_8^0+\frac{l}{1+l}f_6^{'0}+\frac{2\gamma-l}{1+l}f_8^{'0}
    +\frac{2}{1+l}\omega_6\rho\frac{\bm{c}_6\cdot\bm{u}_b}{c_s^2}\right]+(1-r)f_6^{(\text{eq})}(\rho,u_w),
  \end{equation}
\end{subequations}
where $u_w$ is the wall velocity, and $f_i^0=f_i(y_0)$ and $f_i^{'0}=f_i'(y_0)$ with $y_0=\gamma\delta_x$.

Following the procedures exhibited in Refs. \cite{Guo08,WangL20,Guo08P}, we can obtain the relationship between the velocities $u_0$ and $u_1$ respectively at $j=0$ and $j=1$:
\begin{equation}\label{NumerSana}
   u_1=\mathcal{A}u_0+\mathcal{B} a \delta_t+\mathcal{C} u_w,
\end{equation}
where
\begin{subequations}\label{u1u0Coeff}
\begin{equation}\label{CoeAB}
   \mathcal{A}=\frac{(1+l)(1-r)\tau_s+2r(1+\gamma)}{(1+l)(1-r)\tau_s+2r\gamma}, \qquad \mathcal{C}=-\frac{2r}{(1+l)(1-r)\tau_s+2r\gamma},
\end{equation}
\begin{equation} \label{CoeC}
  \mathcal{B}=\frac{(1+r+l-lr)(1+4\tau_q-8\tau_s-8\tau_q\tau_s)
  +12r[(2\gamma-l)(2\tau_s-1)+(1-2\gamma)\tau_s]}{2\left[(1+l)(1-r)\tau_s+2r\gamma\right](2\tau_s-1)}.
\end{equation}
\end{subequations}
For the Poiseuille flow between two plates located at $y=0$ and $y=H$, the analytical solution can be expressed as
\begin{equation}\label{OrSnumve}
   u_j=4u_c\frac{y_j}{H}\left(1-\frac{y_j}{H}\right)+u_w+u_s,
\end{equation}
where $u_c=aH^2/8\nu$, and $u_s$ is the slip velocity at the physical wall. Substituting Eq. \eqref{OrSnumve} into Eq. \eqref{NumerSana} to replace $u_0$ and $u_1$, we can derive the dimensionless slip velocity
\begin{equation}\label{EqMiSlip}
   U_s:=\frac{u_s}{u_c}=\frac{2(1+l)(1-r)}{r}\tau_s\frac{\delta_x}{H}
   -\frac{1}{3r}\left[(1+l)(1+4\tau_q-2\tau_s+12\gamma\tau_s-8\tau_s\tau_q)+r\mathcal{D}\right]\frac{\delta_x^2}{H^2}.
\end{equation}
where $\mathcal{D}=12\gamma^2+12\gamma(\tau_s-1)+(2\tau_s-1)(4l\tau_q-1-4\tau_q)-l[11(2\tau_s-1)+12\gamma\tau_s]$. Further invoking the $\tau_s$-$\text{Kn}$ relation given by Eq. \eqref{tKnrela}, Equation \eqref{EqMiSlip} can be rewritten in terms of $\text{Kn}$ as
\begin{equation}\label{NumeriSlip}
   U_s=4\frac{(1-r)(1+l)\tau_s}{r(2\tau_s-1)}\sqrt{\frac{6}{\pi}}\text{Kn}
   +\frac{8}{\pi}\frac{\mathcal{K}}{r(2\tau_s-1)^2}\text{Kn}^2
\end{equation}
where $\mathcal{K}=(2\tau_s-1)(1+r+l+11lr)+12\gamma[r+\tau_s(lr-l-r-1)]-12r\gamma^2-4\tau_q(2\tau_s-1)(lr-l-r-1)$. The expression of Eq. \eqref{NumeriSlip} is the slip velocity derived from the boundary scheme \eqref{EqNewBDSC}. Compared with previous similar studies \cite{Szal06,Tao15}, the most striking feature here is that the slip velocity $U_s$ includes an additional free parameter $l$ besides $\gamma$. Owing to the degree of freedom from $l$, we now could obtain invariant relaxation parameters to realize the adopted slip boundary condition by the boundary scheme \eqref{EqNewBDSC}.

For non-continuum gaseous flows, a second-order slip boundary condition is widely used and read as \cite{Suga13,Hadji06}
\begin{equation}\label{PhyscSlip}
  u_s=\mathcal{L}_1\lambda\frac{\partial u}{\partial n}-\mathcal{L}_2\lambda^2\frac{\partial^2u}{\partial n^2},
\end{equation}
where $\bm{n}$ is the unit outer vector normal to the wall, and $\mathcal{L}_1$ and $\mathcal{L}_2$ refer to the slippage coefficients related with the gas-wall interactions. Under the slip boundary condition \eqref{PhyscSlip}, the Poiseuille flow exhibits the following slip velocity at the wall
\begin{equation}\label{PhyscPoiSlip}
  U_s=4\mathcal{L}_1\text{Kn}+8\mathcal{L}_2\text{Kn}^2.
\end{equation}
Therefore, to realize the slip boundary condition \eqref{PhyscPoiSlip} with the boundary scheme \eqref{EqNewBDSC}, $r$ and $\tau_q$ must be determined by comparing Eq. \eqref{NumeriSlip} with Eq. \eqref{PhyscPoiSlip}, which are given by
\begin{align}
  r&=\frac{\sqrt{\frac{6}{\pi}}\tau_s(1+l)}{\mathcal{L}_1(2\tau_s-1)+\sqrt{\frac{6}{\pi}}\tau_s(1+l)},\label{SliCoeffrtsr}\\
\tau_q&=\frac{\mathcal{L}_1(2\tau_s-1)\left[2\tau_s(6\gamma-1)+1\right]
+\sqrt{\frac{6}{\pi}}\tau_s\bigl[\mathcal{L}_2\pi(2\tau_s-1)^2+12\gamma^2+2(2\tau_s-1)(6\gamma-6l-1)\bigr]}
{4(2\tau_s-1)\left[2\sqrt{\frac{6}{\pi}}\tau_s+\mathcal{L}_1(2\tau_s-1)\right]}. \label{SliCoeffrtstq}
\end{align}

As noted previously, the parameter $l$ is related to the distance ratio $\gamma$. Then, it can be seen from Eqs. \eqref{SliCoeffrtsr} and \eqref{SliCoeffrtstq} that $r$ and $\tau_q$ are dependent on $\gamma$ as well as the gas-wall interaction parameters and the relaxation time $\tau_s$. For microgaseous flows with curved walls, it is noted that different boundary nodes bring variable values of $\gamma$ along different lattice directions. It is reasonable for the parameter $r$, which represents the fraction of the no-slip velocity condition in the combination, to change locally with boundary nodes. However, for previous curved boundary schemes which only contains $\gamma$, the relaxation time $\tau_q$ therein is inevitably changeable with $\gamma$ to realize the slip boundary condition. In the following, we will fix the choice of $l$ to address the issue of nonuniform relaxation parameters.

From Eq. \eqref{tKnrela}, the relaxation time $\tau_s$ is determined by $\text{Kn}$, and should be constant for a concrete flow problem. Noting that $l$ is a function of $\gamma$, we denote $P(\gamma)=\mathcal{L}_1(2\tau_s-1)\left[2\tau_s(6\gamma-1)+1\right]
+\sqrt{\frac{6}{\pi}}\tau_s\bigl[\mathcal{L}_2\pi(2\tau_s-1)^2+12\gamma^2+2(2\tau_s-1)(6\gamma-6l-1)\bigr]$. In order to achieve the uniform $\tau_q$, the numerator of Eq. \eqref{SliCoeffrtstq} must be irrelevant to $\gamma$, which leads to
\begin{equation}
  \frac{\text{d} P(\gamma)}{\text{d} \gamma}=0 \Rightarrow (1-2\tau_s)\sqrt{\frac{6}{\pi}}\frac{\text{d} l}{\text{d} \gamma}
  +2\sqrt{\frac{6}{\pi}}\gamma +\bigl(\mathcal{L}_1+\sqrt{\frac{6}{\pi}}\bigr)(2\tau_s-1)=0.
\end{equation}
For the above ordinary differential equation, the solution of $l$ is
\begin{equation}\label{Eqexlva}
   l=\frac{\sqrt{\frac{6}{\pi}}\gamma^2+\gamma(2\tau_s-1)\bigl(\mathcal{L}_1+\sqrt{\frac{6}{\pi}}\bigr)}{\sqrt{\frac{6}{\pi}}(2\tau_s-1)}
   +\mathcal{E},
\end{equation}
where $\mathcal{E}$ is a constant independent of $\gamma$. With the solved $l$ substituted into Eqs. \eqref{SliCoeffrtsr} and \eqref{SliCoeffrtstq}, we can obtain the new expressions of $r$ and $\tau_s$ as
\begin{align}
  r&=\frac{\mathcal{L}_1\tau_s\gamma(2\tau_s-1)
  +\sqrt{\frac{6}{\pi}}\tau_s[\gamma^2+\gamma(2\tau_s-1)+(1+\mathcal{E})(2\tau_s-1)]}{\mathcal{L}_1(2\tau_s-1)[(2+\gamma)\tau_s-1]
  +\sqrt{\frac{6}{\pi}}\tau_s[\gamma^2+\gamma(2\tau_s-1)+(1+\mathcal{E})(2\tau_s-1)]},\label{NSliCoertsr}\\
\tau_q&=\frac{(2\tau_s-1)\left(\tau_s\sqrt{6\pi}\mathcal{L}_2-\mathcal{L}_1\right)
-2\sqrt{\frac{6}{\pi}}\tau_s(1+6\mathcal{E})}
{8\sqrt{\frac{6}{\pi}}\tau_s+4\mathcal{L}_1(2\tau_s-1)}. \label{NSliCoertstq}
\end{align}
It is clear that the relaxation time $\tau_q$ is irrelevant to $\gamma$ now. In summary, our approach to resolve the uniform relaxation time $\tau_q$ is prescribed as follow: When $l$ is given by Eq. \eqref{Eqexlva}, the relaxation time $\tau_q$ can take uniform value as Eq. \eqref{NSliCoertstq}, and simultaneously the slip boundary condition \eqref{PhyscSlip} can be correctly realized by the boundary scheme \eqref{EqNewBDSC} with $r$ determined by Eq. \eqref{NSliCoertsr}.

We note that the above treatment cannot be accomplished in previous studies for curved slip walls. For the sake of clarification, we refer to the boundary scheme \eqref{EqNewBDSC} at $l=0$ temporarily as one example of previous curved boundary conditions. Correspondingly, the values of $r$ and $\tau_q$ expressed in Eqs. \eqref{SliCoeffrtsr} and \eqref{SliCoeffrtstq} turn to
\begin{align}
r&=\frac{\sqrt{\frac{6}{\pi}}\tau_s}{\mathcal{L}_1(2\tau_s-1)+\sqrt{\frac{6}{\pi}}\tau_s},\label{SliCoeffrtsrD}\\
\tau_q&=\frac{\mathcal{L}_1(2\tau_s-1)\left[2\tau_s(6\gamma-1)+1\right]
+\sqrt{\frac{6}{\pi}}\tau_s\bigl[\mathcal{L}_2\pi(2\tau_s-1)^2+12\gamma^2+2(2\tau_s-1)(6\gamma-1)\bigr]}
{4(2\tau_s-1)\left[2\sqrt{\frac{6}{\pi}}\tau_s+\mathcal{L}_1(2\tau_s-1)\right]}. \label{SliCoeffrtstqrD}
\end{align}
This clearly indicates that without the parameter $l$ in previous studies, the relaxation time $\tau_q$ should be changeable with $\gamma$ to realize the prescribed slip boundary condition, as noted before. For the case of flat walls, $\tau_q$ can be invariant because the distance ratio is fixed for all boundary nodes. While for the case of curved wall geometries, approximated measurements for $\gamma$ would not be avoided to obtain the uniform $\tau_q$. However, as will shown later in the numerical examples, this hinders us to correctly realize the slip boundary condition \eqref{PhyscSlip} \cite{Silva17}. Specifically, corresponding to the halfway DBB scheme, Eqs. \eqref{SliCoeffrtsrD} and \eqref{SliCoeffrtstqrD} at $\gamma=1/2$ give the parameters of $r$ and $\tau_q$ that are required to realize the prescribed slip boundary condition
\begin{align}
r&=\frac{\sqrt{\frac{6}{\pi}}\tau_s}{\mathcal{L}_1(2\tau_s-1)+\sqrt{\frac{6}{\pi}}\tau_s},\label{SliCoeffHar}\\
\tau_q&=\frac{\mathcal{L}_1(2\tau_s-1)\left(4\tau_s+1\right)
+\sqrt{\frac{6}{\pi}}\tau_s\bigl[\mathcal{L}_2\pi(2\tau_s-1)^2+4(2\tau_s-1)+3\bigr]}
{4(2\tau_s-1)\left[2\sqrt{\frac{6}{\pi}}\tau_s+\mathcal{L}_1(2\tau_s-1)\right]}. \label{SliCoeffHata}
\end{align}

Now some comments on the above derivation results are given in order. First, if we set $\mathcal{L}_1=\mathcal{L}_2=0$ in Eq. \eqref{PhyscSlip}, the slip velocity $u_s$ degenerates to $u_s=0$, which means the no-slip boundary condition at solid walls. As a consequence, the combination parameter $r$ in Eq. \eqref{NSliCoertsr} becomes to $r=1$, and the boundary scheme \eqref{EqNewBDSC} reduces to that for the no-slip boundary condition \cite{Zhao19}, as pointed out in Remark \ref{r1Rem}. Then, the values of $l$ and $\tau_q$ in Eqs. \eqref{Eqexlva} and \eqref{NSliCoertstq} will follow the same method as those given in our recent work \cite{WangL20} for curved no-slip walls. Second, to treat curved slip walls with the uniform $\tau_q$, previous studies usually approximate the distance ratio $\gamma$ as $\gamma=\frac{1}{2}$ (i.e., the halfway boundary scheme) or by an artificial measurement. These inaccurate values of $\gamma$ degrade the fidelity of curved geometry. Further based on Eq. \eqref{SliCoeffrtstqrD}, the actual $\tau_q$ cannot be obtained to realize the slip boundary condition under coarse grid resolutions. Third, although the above derivations are based on the planar Poiseuille flow, they may be applicable to general cases with curved walls if the flows in the near-wall region can be assumed to have a second-order polynomial profile locally. This assumption has been adopted in many previous studies, and its reasonability has been demonstrated for continuum flows and microscale gaseous flows \cite{Verhaeghe09,Chai08,Chai08,Guo07,Guo08,Tao15,Guo11,Silva17,WangL20,He97,Ginzburg03,Pan06}. Finally, there are many choices of $l$ to derive infinitely specific curved boundary conditions from Eq. \eqref{EqNewBDSC}. The present theoretical analysis clearly indicates that in order to correctly realize a certain slip boundary condition at curved walls, the free parameter $l$ besides $r$ and $\tau_q$ must be also chosen carefully to ensure invariable relaxation parameters. Eqs. \eqref{Eqexlva}-\eqref{NSliCoertstq} give the theoretical formulae to determine $l$ and $r$ in the boundary scheme \eqref{EqNewBDSC} together with the unform $\tau_q$.

Within the framework of BGK model, the corresponding $l$, $r$ and $\tau_q$ in the above equations can be determined by taking $\tau_s=\tau_q=\tau$. After a direct comparison, one can find that the relaxation time $\tau$ from Eq. \eqref{NSliCoertstq} remains invariant to $\gamma$, however, $\tau$ from Eq. \eqref{SliCoeffrtstqrD} without $l$ still changes with $\gamma$ as its MRT counterpart. Based on this fact, we can conclude that only by seeking more relaxation parameters in the LBE (e.g., extending the BGK model to the MRT model), the uniform relaxation time cannot be accomplished to realize the slip boundary condition at curved walls. To resolve such problem, one more feasible method based on our analysis is by adding free parameters to the curved boundary scheme.

Finally, we discuss the parameter range in Eqs. \eqref{Eqexlva}-\eqref{NSliCoertstq} adapted to the curved boundary scheme \eqref{EqNewBDSC}. To ensure the non-negativity of $l$ in Eq. \eqref{Eqexlva}, the constant $\mathcal{E}$ should satisfy
\begin{equation}
   \mathcal{E}\geq-\frac{\sqrt{\frac{6}{\pi}}\gamma^2+\gamma(2\tau_s-1)\bigl(\mathcal{L}_1
   +\sqrt{\frac{6}{\pi}}\bigr)}{\sqrt{\frac{6}{\pi}}(2\tau_s-1)}.
\end{equation}
With the above requirement together with $\tau_s>0.5$, one can derive that the numerator and denominator in Eq. \eqref{NSliCoertsr} are both positive, and further the combination parameter $r$ locates in $0<r<1$. On the other hand, because of the stability condition that $\tau_q>0.5$, another requirement of $\mathcal{E}$ from Eq. \eqref{NSliCoertstq} can be obtained as
\begin{equation}\label{EqECond}
    \mathcal{E}<\frac{(\pi\tau_s\mathcal{L}_2-3\sqrt{\frac{\pi}{6}}\mathcal{L}_1)(2\tau_s-1)}{12\tau_s}-\frac{1}{2}.
\end{equation}
For the case that $\gamma$ is close to zero, the parameter $l$ from Eq. \eqref{Eqexlva} approximates to $\mathcal{E}$ and hence should obey Eq. \eqref{EqECond} as well. When the right-hand-side term is smaller than zero, the parameter $l$ in the simulations would be negative. In this case, as we have done for continuum flows \cite{WangL20}, to balance the numerical stability and the accurate implementation of the boundary scheme \eqref{EqNewBDSC}, the parameter $l$ is compulsively set as zero if the actual $l<-0.4$ obtained from Eq. \eqref{Eqexlva} at very small $\gamma$. As will be shown in the subsequent simulations, this compromised treatment can bring good predictions with the analytical solutions.

\section{Numerical results and discussions}\label{Sec4}
To validate the proposed curved boundary condition [Eq. \eqref{EqNewBDSC}] and the theoretical derivations, some well-established microscale flows are simulated in this section. The problems under consideration include the microscale aligned and inclined plane Poiseuille flow and the microcylindrical Couette flow. The slip boundary condition at curved walls is modeled by the boundary scheme \eqref{EqNewBDSC}, where $l$ and $r$ are determined from Eqs. \eqref{Eqexlva} and \eqref{NSliCoertsr}. For comparison with previous studies, some other choices of $l$ as well as the halfway boundary scheme (corresponds to $\gamma=1/2$) are also tested to predict the microslip velocity at curved walls. To clearly expose the difference between them, coarse grid resolutions are used in the numerical simulations of all considered problems.

In the simulations, the slip coefficients $\mathcal{L}_1$ and $\mathcal{L}_2$ in Eq. \eqref{PhyscSlip} are taken as $\mathcal{L}_1=(2-\sigma)(1-0.1817\sigma)/\sigma$, and $\mathcal{L}_2=\pi^{-1}+\mathcal{L}_1^2/2$ \cite{Guo08P}, where $\sigma$ is the wall accommodation coefficient. The value of $\sigma$ in this work is set as $\sigma=1$, which means that the wall is fully diffusive. The relaxation times $\tau_\rho$ and $\tau_j$ are specified as $\tau_\rho=\tau_j=1.0$, and $\tau_s$ is determined via the Knudsen number as Eq. \eqref{tKnrela}. To realize the slip boundary condition by the proposed boundary scheme \eqref{EqNewBDSC}, $\tau_q$ is chosen according to Eq. \eqref{NSliCoertstq} as derived above. The remaining relaxation times $\tau_e$ and $\tau_\varepsilon$ are given by $\tau_e=1.1$ and $\tau_\varepsilon=1.2$. Actually, the effects of $\tau_\rho$, $\tau_j$, $\tau_e$ and $\tau_\varepsilon$ are negligible on the numerical results. For the diffusive scaling used in the boundary scheme \eqref{EqNewBDSC}, $\delta_x$ and $\delta_t$ obey the relation of $\delta_t=\eta\delta_x^2$, and in the simulations, they are determined as $\delta_t=\eta\delta_x^2, ~\eta=\frac{\tau_s-\frac{1}{2}}{3\nu}$.

\subsection{Force-driven microchannel flow}
We first consider the Poiseuille flow driven by a constant force in a microchannel. This microflow, which has an analytical solution, has been recognized as a benchmark problem in the LBM. In the following, the aligned and inclined plates with respect to the computational grid are successively considered in the simulations.

\subsubsection{Aligned channel case}
\begin{figure}
\centering
\includegraphics[width=0.55\textwidth]{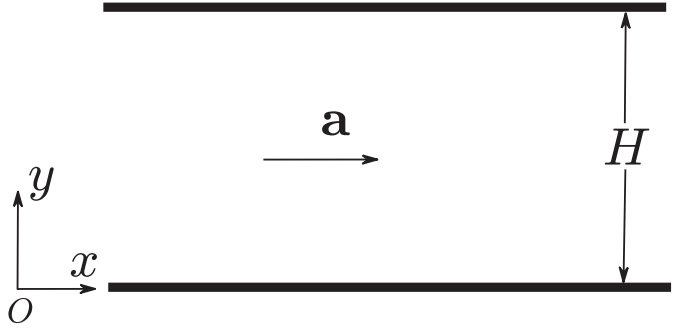}
\caption{Schematic of the Poiseuille flow in a microscale channel with aligned straight walls.}
\label{ScheAlPois}
\end{figure}

The force-driven Poiseuille flow between two parallel plates is first simulated. As schematically shown in Fig. \ref{ScheAlPois}, the flows in the channel with width $H$ are driven by a constant acceleration $\bm{a}=(a,0)$ along the $x$-direction.
Periodic boundary conditions are executed at the entrance and outlet of the channel. With the wall slip velocity given by Eq. \eqref{PhyscPoiSlip}, the dimensionless velocity has the following analytical solution
\begin{equation}\label{numvelo}
   U:=\frac{u(y)}{u_c}=4\frac{y}{H}\left(1-\frac{y}{H}\right)+4\mathcal{L}_1\text{Kn}+8\mathcal{L}_2\text{Kn}^2,\quad  V:=\frac{v(y)}{u_c}=0,
\end{equation}
where $0\leq y \leq H$, and $u_c=aH^2/8\nu$ is the maximum streamwise velocity.

In the simulations, the lower and upper plates are placed with distance $\gamma \delta_x$ away from boundary lattice nodes (as illustrated in Fig. \ref{ScheMiclattice}). The grid number spanning in the vertical direction is $M$, and this gives the lattice spacing as $\delta_x=H/(M+2\gamma)$. To ensure the low Mach number for a finite Knudsen number, the driven acceleration $a$ is set to be $10^{-4}$. We performed some simulations with the boundary scheme \eqref{EqNewBDSC} to measure the dimensionless slip velocities under different choices of $l$.
\begin{figure}
\begin{tabular}{cc}
\includegraphics[width=0.5\textwidth,height=0.26\textheight]{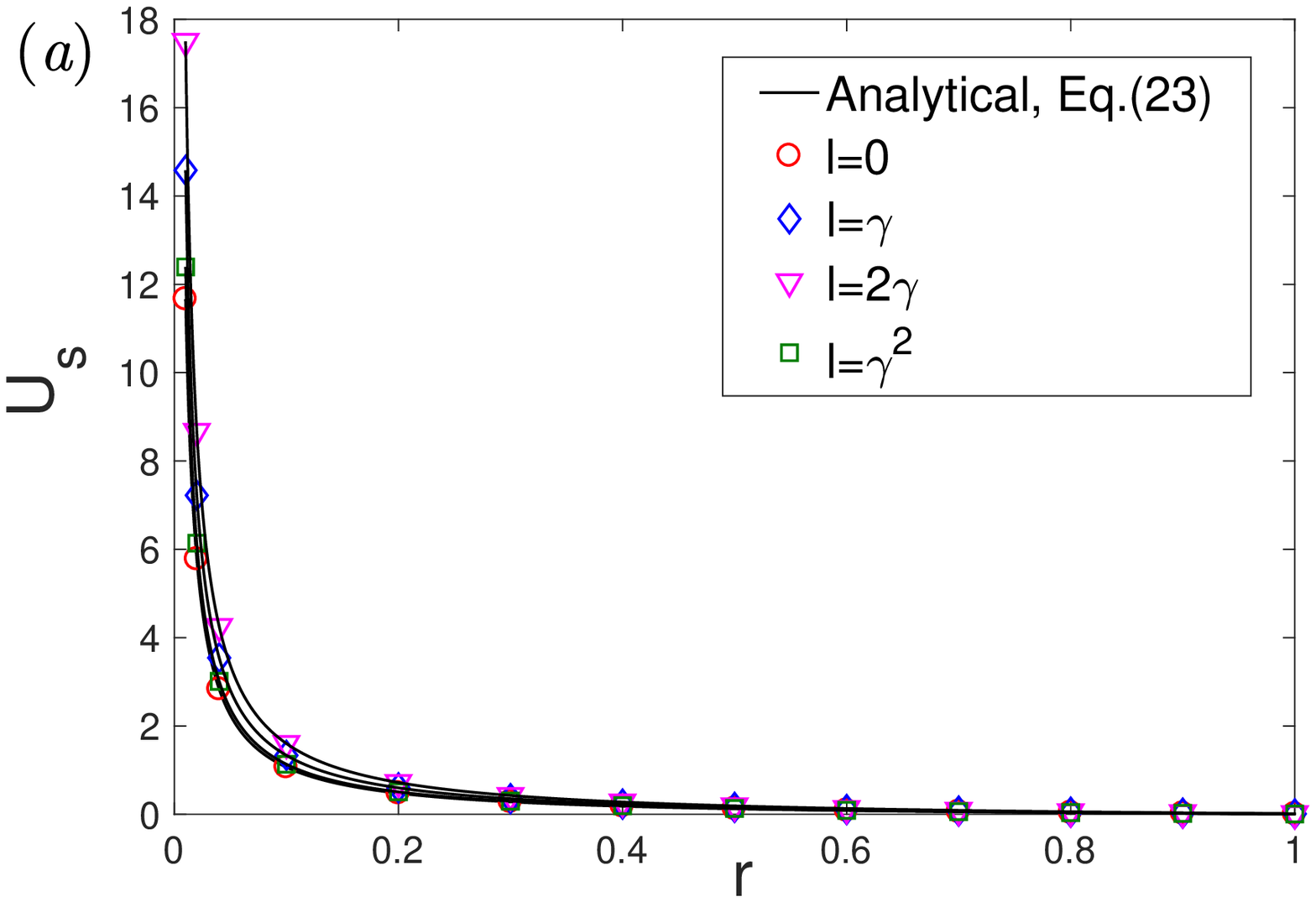}
\includegraphics[width=0.5\textwidth,height=0.26\textheight]{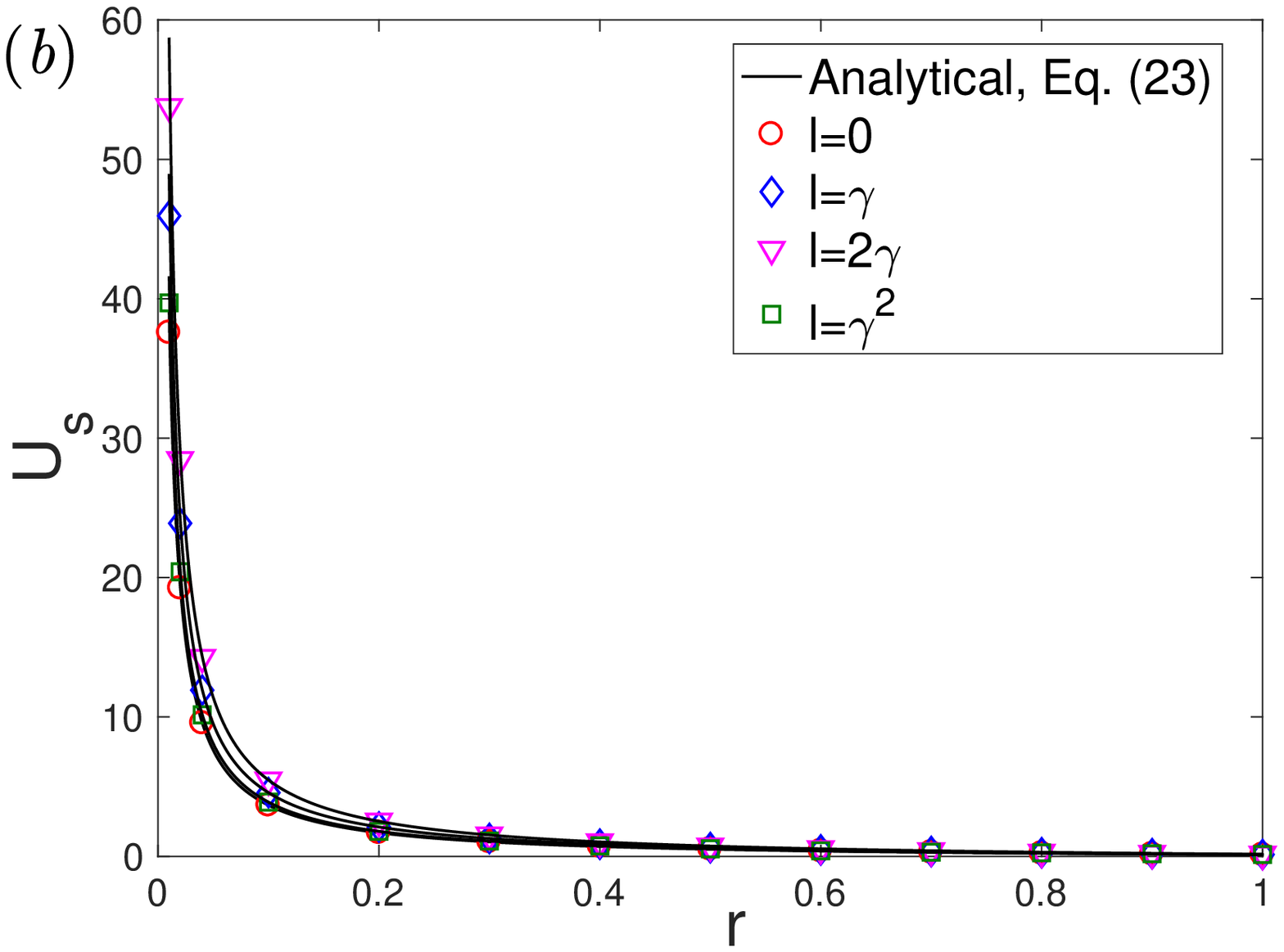}\\
 (a) $\text{Kn}=0.02$  \hspace{20em} (b) $\text{Kn}=0.1$
\end{tabular}
\caption{Slip velocity against $r$ for the boundary scheme \eqref{EqNewBDSC} with different $l$ at $M=16$ and [(a) $\text{Kn}=0.02$; (b) $\text{Kn}=0.1$]. As a specific but representative case, the distance ratio is $\gamma=0.25$, and $\tau_q=\tau_s$.  }
\label{uMSlipPois}
\end{figure}
The numerical results against $r$ at two Knudsen numbers ($\text{Kn}=0.02,~0.1$) are shown in Fig. \ref{uMSlipPois} where $M=16$, $\nu=0.01$, and $\tau_q=\tau_s$. The value of $\gamma=0.25$ is taken as a representative case for $\gamma$ ranging in $0\leq\gamma\leq1$ . As clearly seen from the figure, the numerical predictions agree well with the theoretical derivations given by Eq. \eqref{NumeriSlip}. In addition, similar excellent consistency results can be also obtained for some other values of $\gamma$ and $\tau_q$, which confirms the derivation result for the slip velocity $U_s$.

\begin{figure}
\begin{tabular}{cc}
\includegraphics[width=0.5\textwidth,height=0.26\textheight]{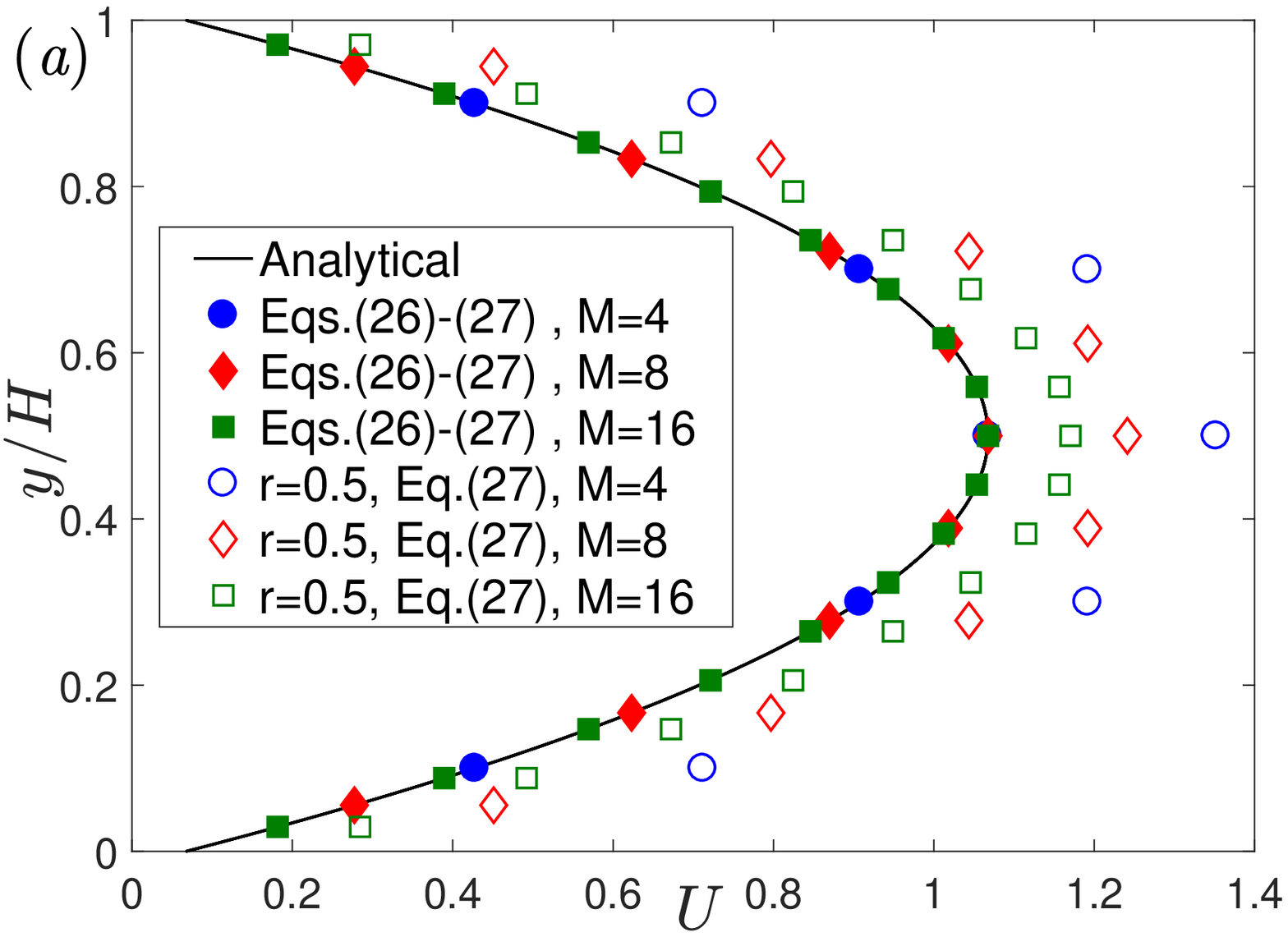}
\includegraphics[width=0.5\textwidth,height=0.26\textheight]{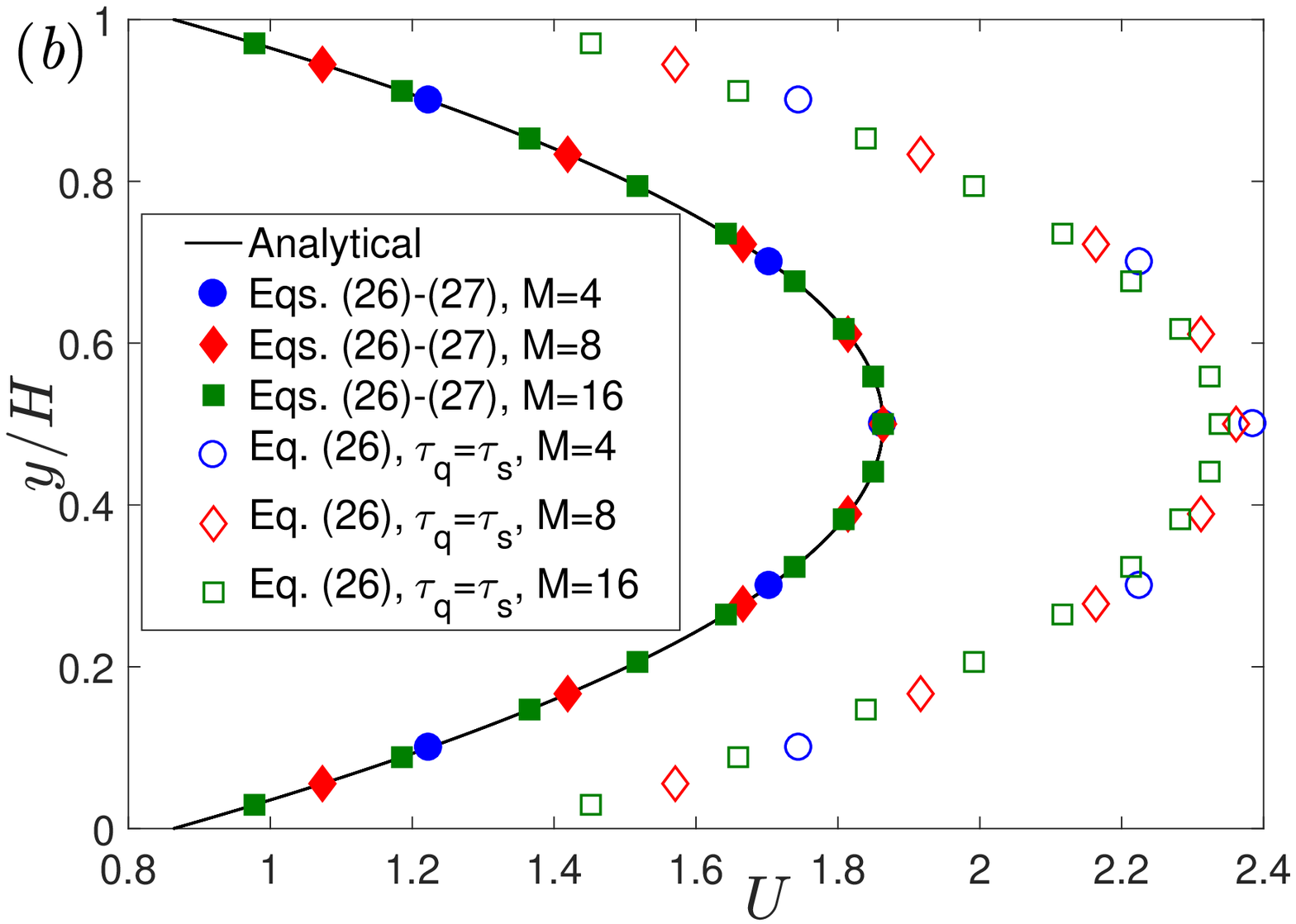}\\
\end{tabular}
\caption{Velocity profiles predicted by the present boundary scheme with different lattice sizes [(a) $\text{Kn}=0.02$; (b) $\text{Kn}=0.2$]. Filled shapes denote the results obtained by Eqs. \eqref{SliCoeffrtsr} and \eqref{SliCoeffrtstq}. Empty shapes denote the results obtained under (a) $r=0.5$ and $\tau_q$ given by Eq. \eqref{SliCoeffrtstq}; (b) $r$ given by Eq. \eqref{SliCoeffrtsr} while $\tau_q=\tau_s$. The analytical velocity solutions (solid lines) are also included.}
\label{uMPoisDise}
\end{figure}

For the case of aligned channel flows, the distance ratios possess the same value at boundary nodes along different lattice directions. Thus, as noted previously, the issue of nonuniform relaxation times will not occur for the halfway DBB and curved boundary schemes to realize the slip boundary condition. We next investigate the discrete effects only for the proposed boundary condition \eqref{EqNewBDSC}. Simulations with different lattice sizes are carried out for $\text{Kn}=0.02$ and $\text{Kn}=0.2$. The predicted velocity profiles are shown in Fig. \ref{uMPoisDise}, where $\gamma=0.5$ and $l=\gamma$ are used as a representative value.
It is clearly seen that if $r$ and $\tau_q$ is unitedly given by Eqs. \eqref{SliCoeffrtsr} and \eqref{SliCoeffrtstq}, the numerical predictions agree well with the analytical solution even with only four grid points. Otherwise, apparent grid-independent results
are observed to deviate from the analytical velocity profiles. Furthermore, as $\text{Kn}$ increases or $\delta_x$ decreases, we found that such difference from the analytical solutions become more pronounced in the simulations. These results demonstrate that the parameters $r$ and $\tau_q$ must be carefully chosen to realize the desired slip boundary condition, as have been revealed in many published studies.

\subsubsection{Inclined channel case}
\begin{figure}
\centering
\includegraphics[height=0.45\textwidth,width=0.7\textwidth]{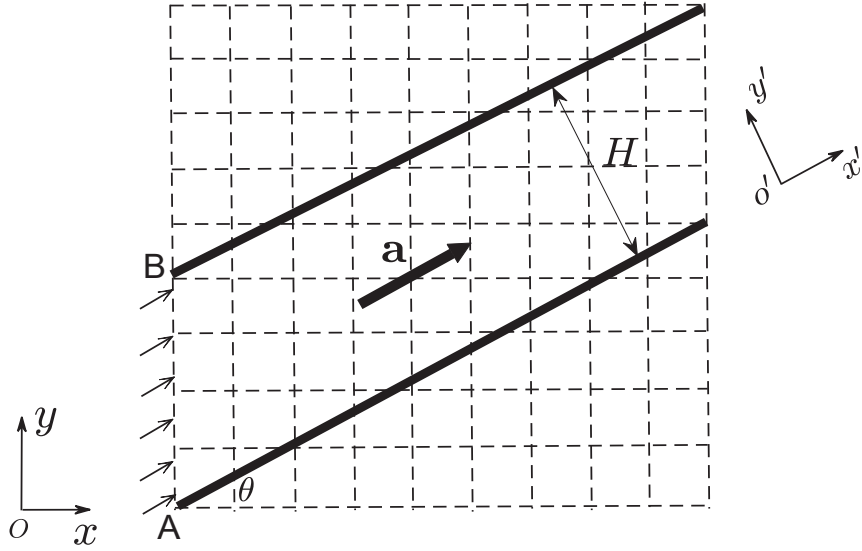}
\caption{Schematic illustration of the Poiseuille flow in an inclined microchannel.}
\label{ScheInPois}
\end{figure}

In contrast to the above aligned microchannel case, the Poiseuille flow in an inclined channel is more complex and further considered.  As shown in Fig. \ref{ScheInPois}, the flat walls are inclined at an inclination angle $\theta$ with respect to the underlying grid.
From the entrance to the exit, the flow in the inclined channel is periodic along the direction of driven force. The microgaseous flows in the channel have the analytical velocity profiles as
\begin{equation}
   U:=\frac{u(y')}{u_c}=4\frac{y'}{H}\left(1-\frac{y'}{H}\right)+4\mathcal{L}_1\text{Kn}+8\mathcal{L}_2\text{Kn}^2,
\end{equation}
where $x'$ and $y'$ are the coordinates respectively parallel and perpendicular to the inclined channel wall, $0\leq y'\leq H $, and $u_c=aH^2/8\nu$.

\begin{figure}
\begin{tabular}{cc}
\includegraphics[width=0.365\textwidth,height=0.20\textheight]{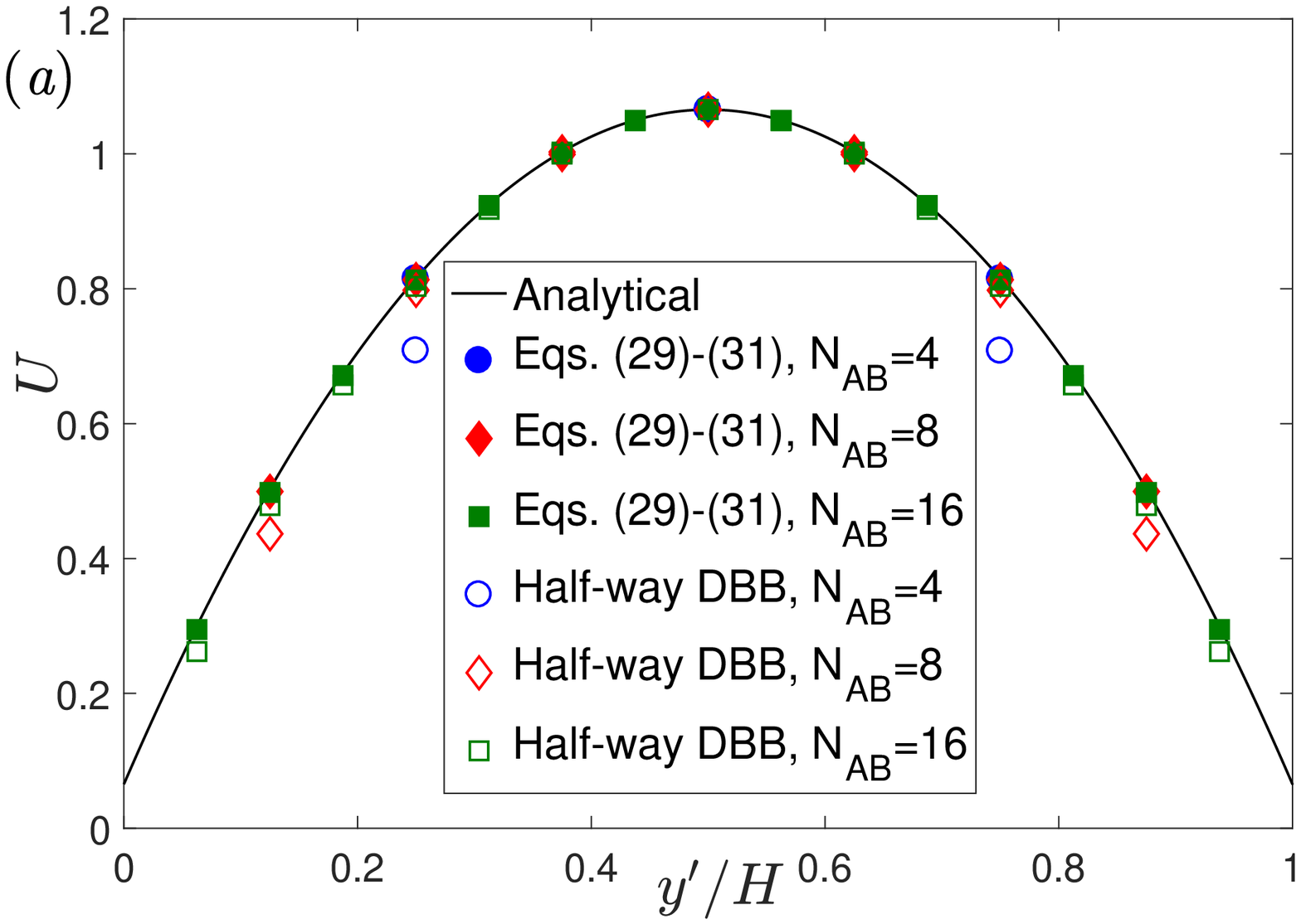}
\includegraphics[width=0.365\textwidth,height=0.20\textheight]{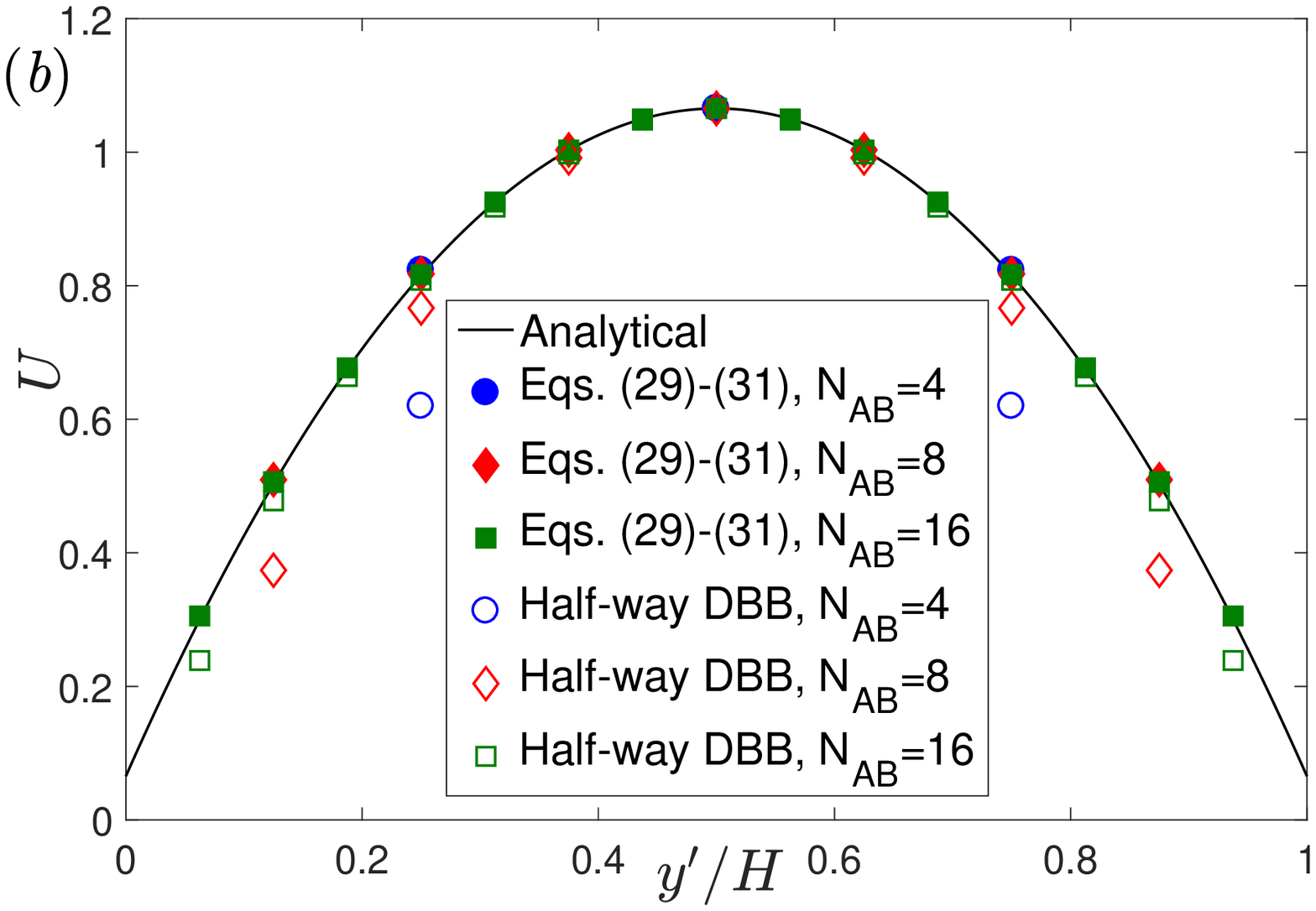}
\includegraphics[width=0.365\textwidth,height=0.20\textheight]{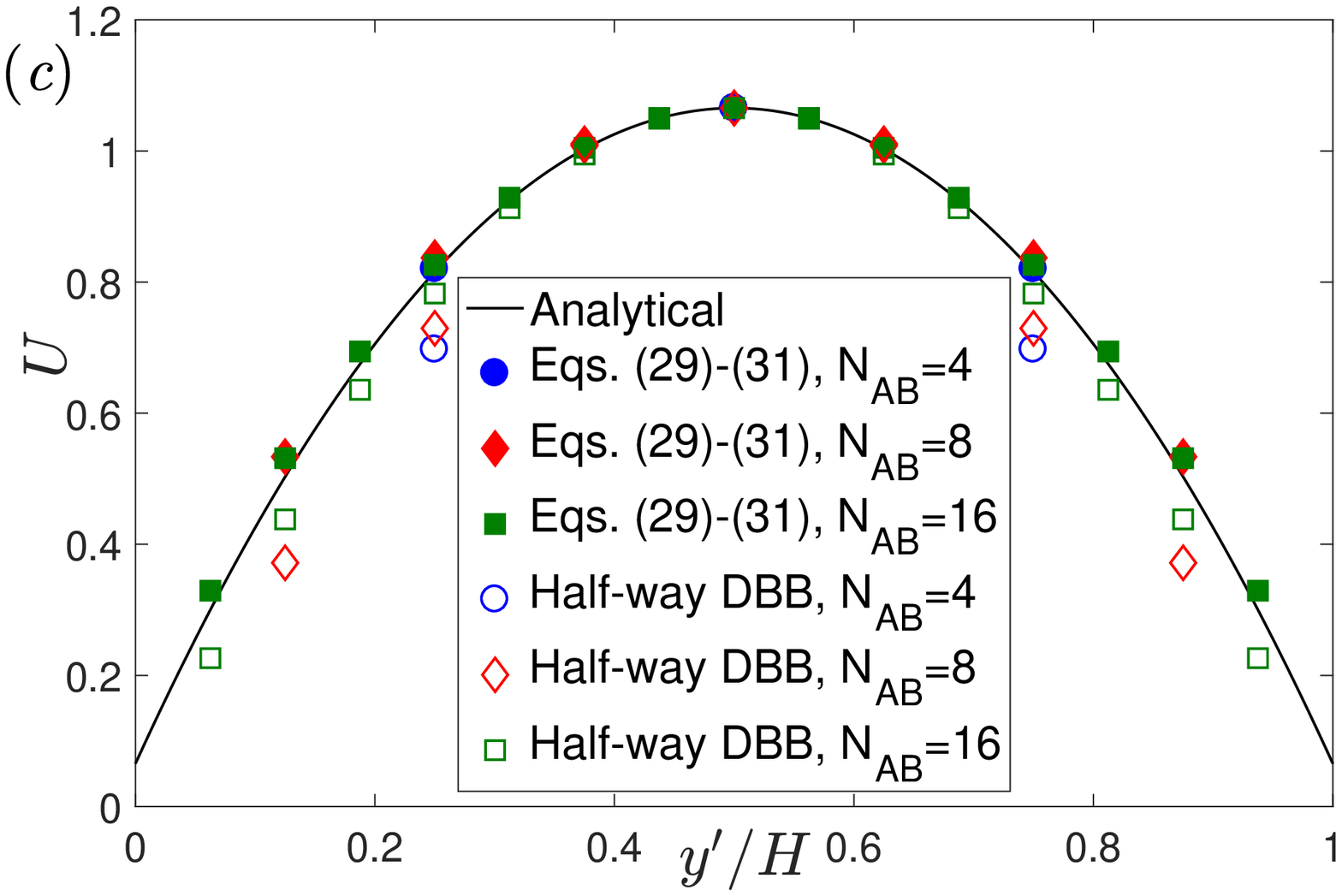}\\
\end{tabular}
\caption{Normalized velocity profiles of the inclined microchannel flow with $\text{Kn}=0.0194$ for different inclination angles [(a) $\text{tan}\theta=0.2$; (b) $\text{tan}\theta=1.2$; (c) $\text{tan}\theta=2.0$]. Filled shapes denote the predicted results by the boundary scheme \eqref{EqNewBDSC} with Eqs. \eqref{Eqexlva}-\eqref{NSliCoertstq}. Empty shapes represent the results obtained by the halfway DBB scheme with Eqs. \eqref{SliCoeffHar} and \eqref{SliCoeffHata}.}
\label{uSliMPois}
\end{figure}

Denote $N_x$ and $N_{AB}$ as the grid number in the $x$-direction and for the segment of $AB$. It follows that the total grid number in the $y$-direction is $N_y=N_{AB}+N_x\text{tan}\theta$. It should be noted that different from the aligned case, the present skew boundary geometry cannot bring uniform values of $\gamma$ at boundary nodes any more. The curved boundary scheme \eqref{EqNewBDSC} with the derived Eqs. \eqref{Eqexlva}-\eqref{NSliCoertstq} under $\mathcal{E}=-0.65$ is then employed for the slip boundary condition at the inclined plates. In the simulations, if the computation from Eq. \eqref{Eqexlva} gives $l<-0.4$ at a boundary node with very small $\gamma$, the parameter $l$ will be set to zero as note before. Figure \ref{uSliMPois} presents the velocity profiles in the inclined microchannel at $\text{Kn}=0.0194$. Three inclination angles, $\text{tan}\theta=0.2,~1.2$, and $2.0$, are considered under $N_x=120$ but different values of $N_{AB}$.
Clearly, good agreement with the analytical solution is achieved for the boundary scheme \eqref{EqNewBDSC} with Eqs. \eqref{Eqexlva}-\eqref{NSliCoertstq}, even with the artificial treatment of $l$ at very small $\gamma$. For comparisons, the results predicted by the halfway DBB scheme with Eqs. \eqref{SliCoeffHar} and \eqref{SliCoeffHata} are also shown. Grid-dependent derivations from the analytical velocity profile are clearly observed. This confirms that notwithstanding the uniform $\tau_q$ determined by Eq. \eqref{SliCoeffHata}, the halfway DBB scheme cannot realize the accurate slip boundary condition at curved walls theoretically. As noted previously, the numerical error is induced by the insufficient accuracy of discrete zigzag ghost boundary to match the real curved wall. In contrast, owing to the local $\gamma$ handling the actual curved geometry and the free parameter $l$ [Eq. \eqref{Eqexlva}], the present boundary scheme \eqref{EqNewBDSC} ($r$ given by Eq. \eqref{NSliCoertsr}) can excellently capture the analytical solution with a uniform $\tau_q$ [Eq. \eqref{NSliCoertstq}] even under a low grid resolution.

\begin{figure}
\begin{tabular}{cc}
\includegraphics[width=0.365\textwidth,height=0.20\textheight]{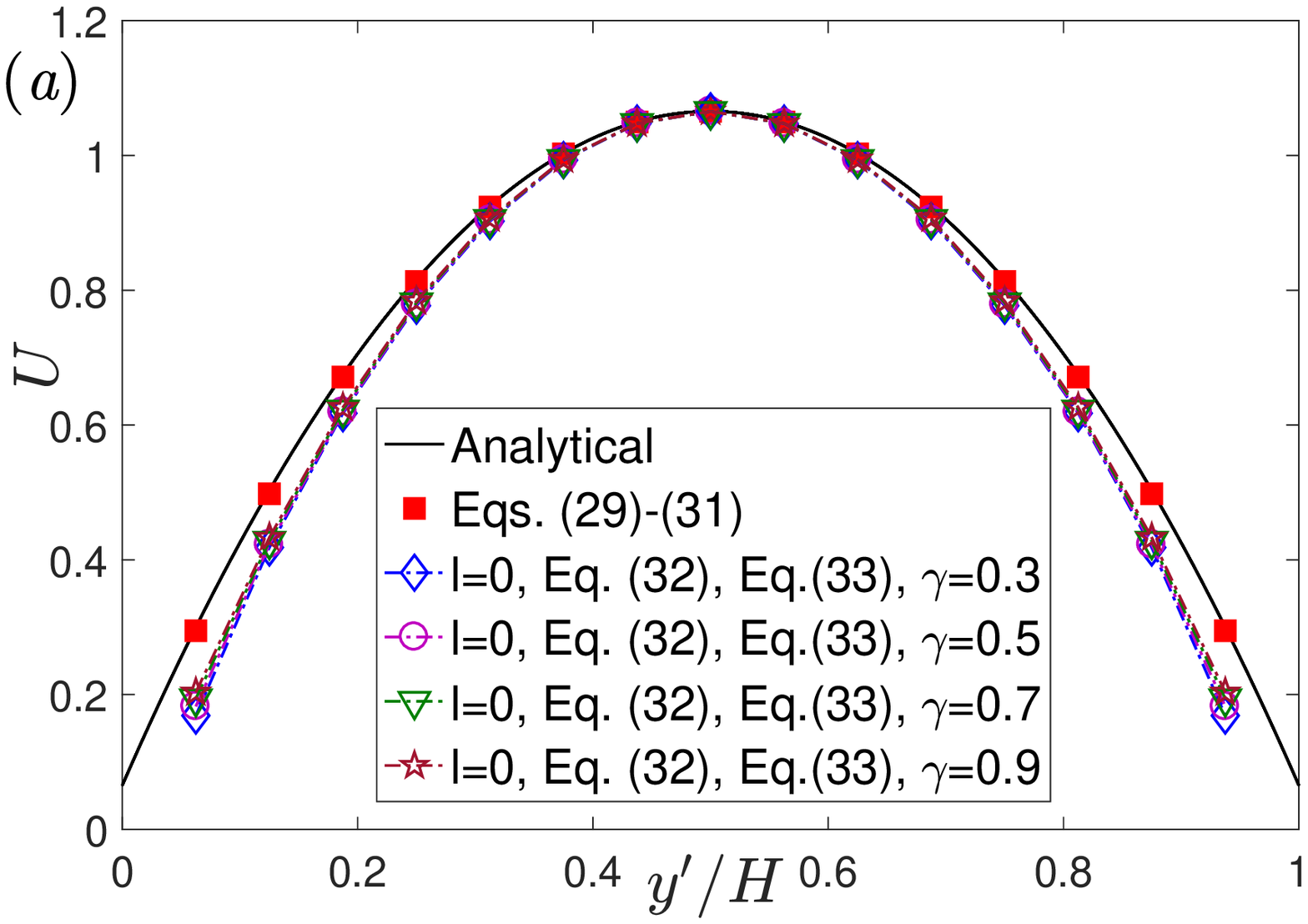}
\includegraphics[width=0.365\textwidth,height=0.20\textheight]{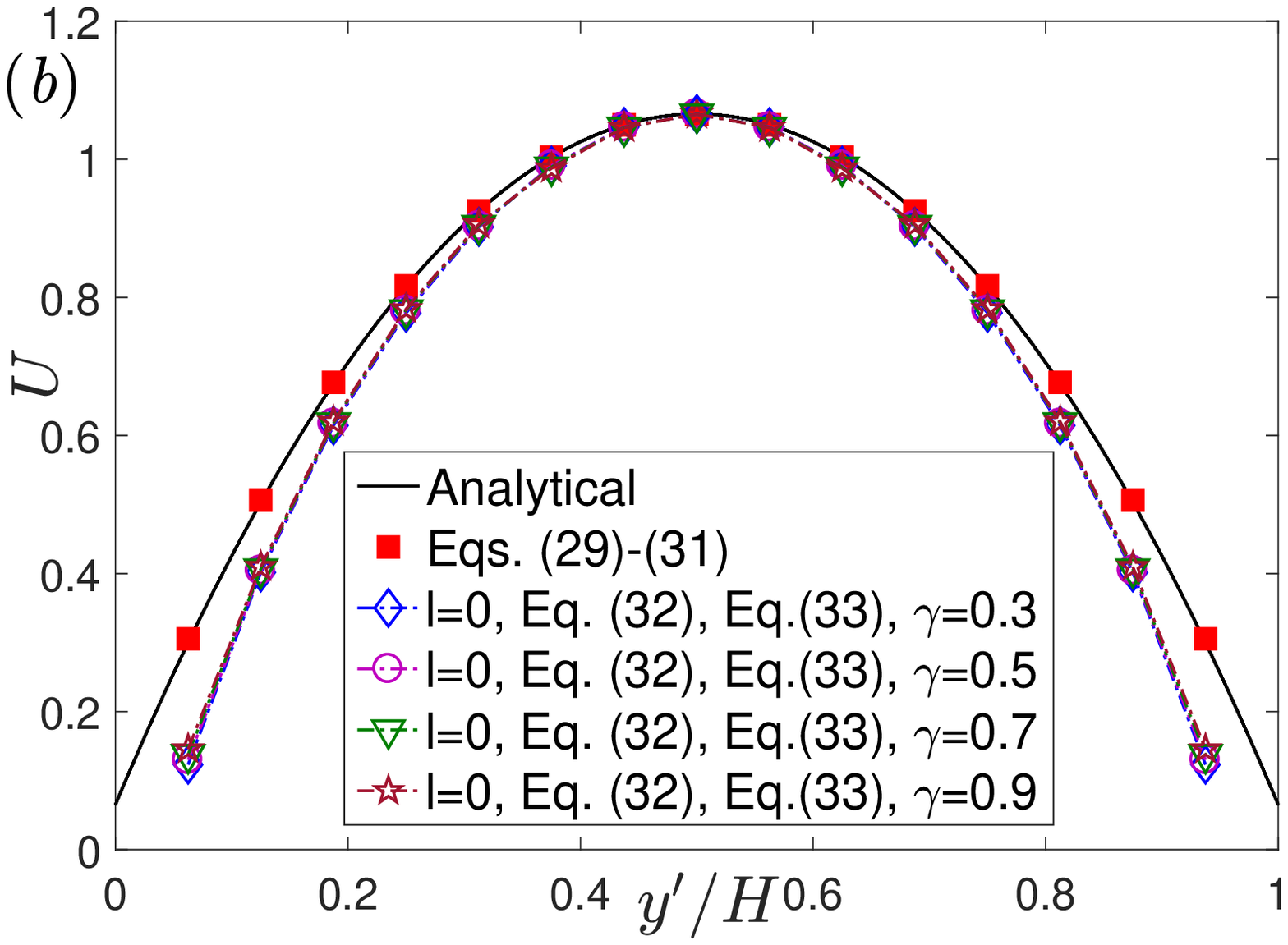}
\includegraphics[width=0.365\textwidth,height=0.20\textheight]{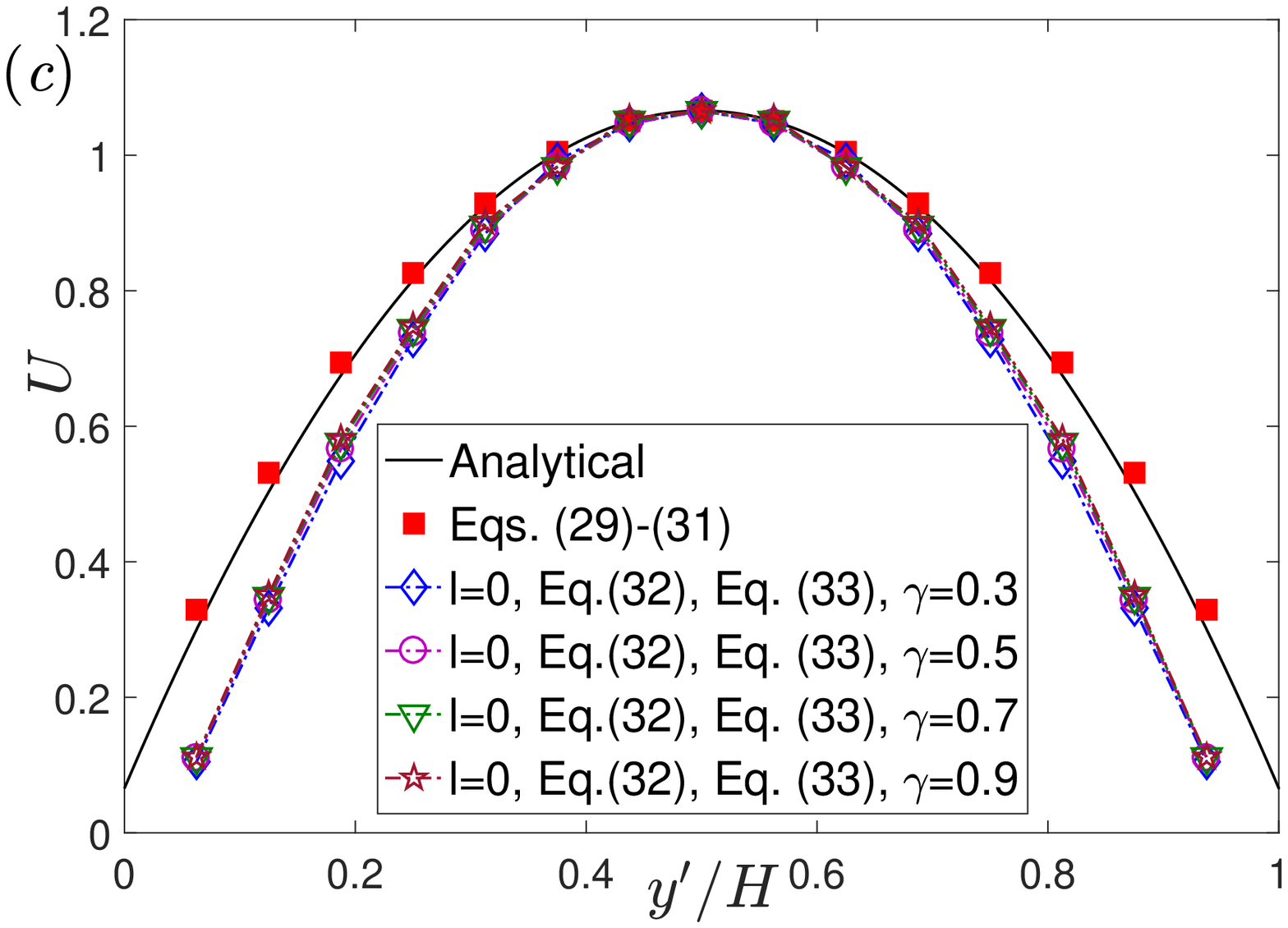}\\
\end{tabular}
\caption{Velocity profiles of the inclined microchannel flow at $\text{Kn}=0.0194$ and $N_{AB}=16$ for different inclination angles [(a) $\text{tan}\theta=0.2$; (b) $\text{tan}\theta=1.2$; (c) $\text{tan}\theta=2.0$]. Filled square shapes denote the predicted results by the boundary scheme \eqref{EqNewBDSC} with Eqs. \eqref{Eqexlva}-\eqref{NSliCoertstq}. Empty shapes represent the results obtained by the boundary scheme \eqref{EqNewBDSC} with $l=0$ and $r$ given by Eq. \eqref{SliCoeffHar}, while $\gamma$ is fixed in Eq. \eqref{SliCoeffHata} to obtain the uniform $\tau_q$.}
\label{uerrSliMPois}
\end{figure}

In Fig. \ref{uerrSliMPois}, the velocity profiles predicted by the boundary scheme \eqref{EqNewBDSC} with $l=0$ and $r$ given by Eq. \eqref{SliCoeffrtsrD} are shown against different $\gamma$ at $\text{Kn}=0.0194$ and $N_{AB}=16$. For each $\gamma$, the relaxation time $\tau_q$ is computed according to Eq. \eqref{SliCoeffrtstqrD}. In this case, this invariable $\tau_q$ is artificially obtained by approximating the same value of $\gamma$ at all boundary nodes. As can be clearly seen, the simulated velocity profiles deviate from the analytical solution in all cases. Furthermore, as also exposed in Fig. \ref{uSliMPois}, such deviations aggravate as the inclination angle $\theta$ increases. Similar deviation results are also found in the simulations for other choices of $l$ that dissatisfy Eq. \eqref{Eqexlva}.
These results demonstrate that for the halfway DBB scheme and previous curved boundary schemes only including $\gamma$, the slip boundary condition at curved walls cannot be successfully realized with uniform relaxation times. In contrast, when $l$ is determined by Eq. \eqref{Eqexlva}, the numerical outcome can produce good predictions consistent with the analytical solution.

From the theoretical analysis presented in Sec. \ref{Sec3}, it is found that the parameter $l$ in the boundary scheme \eqref{EqNewBDSC} affects the numerical slip velocity to match the physical one. To investigate the effect from the choice of $l$, some simulations are carried out with four cases of $l$ ($l=\gamma,~2\gamma,~\gamma^2,~\gamma^2+\gamma$) besides the case of $l$ given by Eq. \eqref{Eqexlva}, while $r$ and $\tau_q$ are fixedly determined by Eqs. \eqref{NSliCoertsr} and \eqref{NSliCoertstq} with $\mathcal{E}=-0.65$.
\begin{figure}
\begin{tabular}{cc}
\includegraphics[width=0.5\textwidth,height=0.25\textheight]{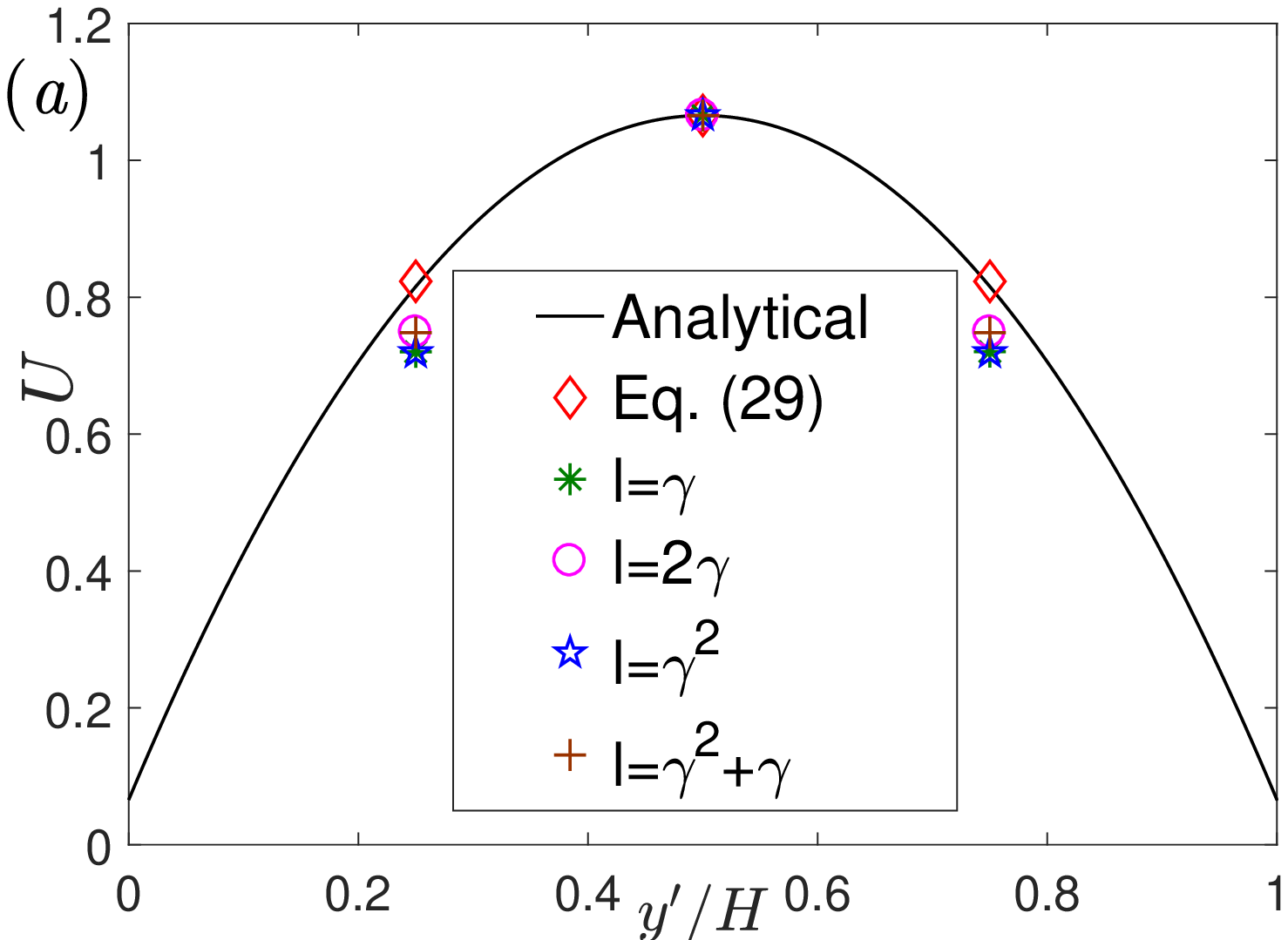}
\includegraphics[width=0.5\textwidth,height=0.25\textheight]{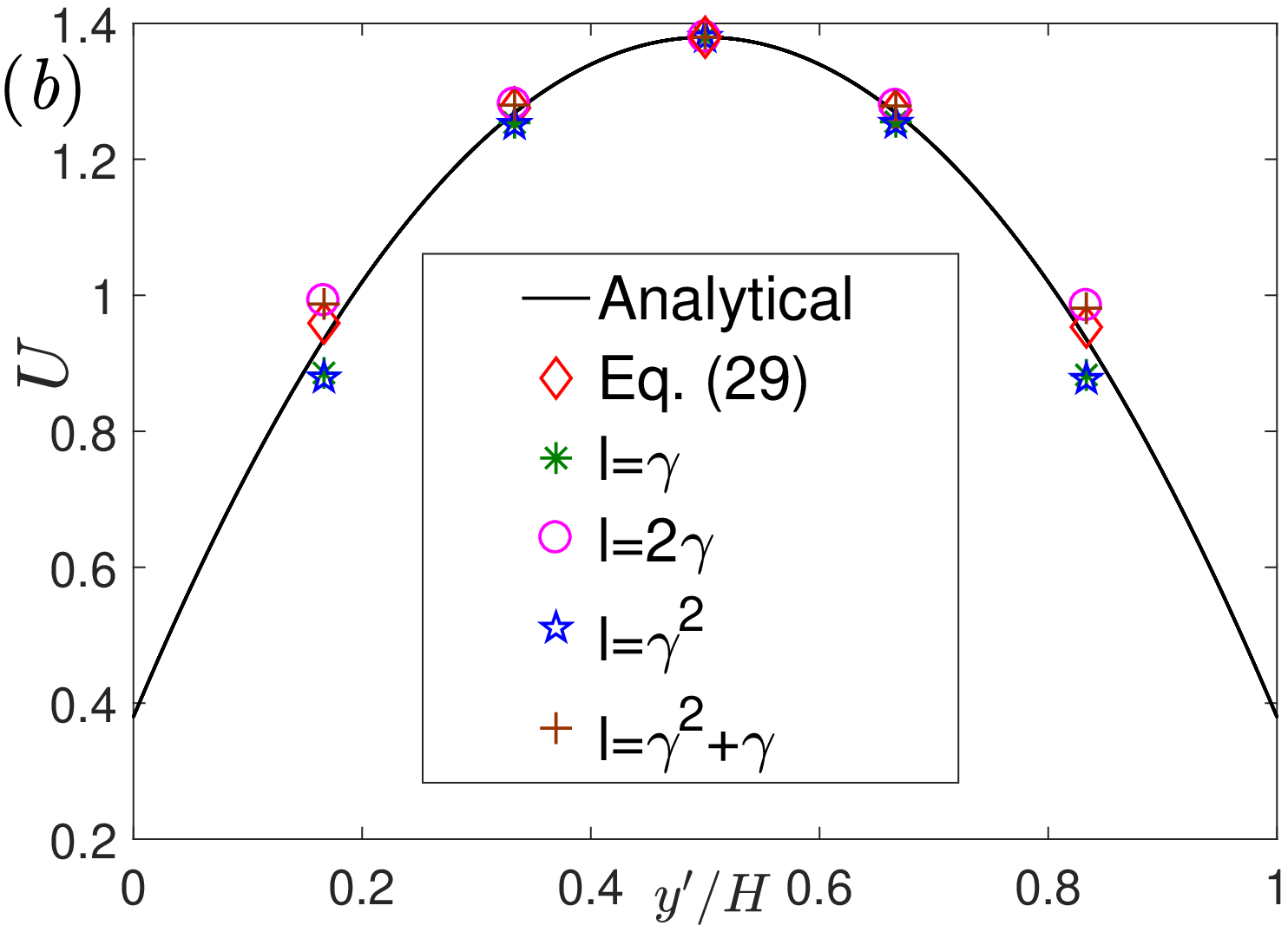}\\
\end{tabular}
\caption{Velocity profiles of the inclined microchannel for different choices of $l$ at $\text{tan}\theta=1.2$. (a) $\text{Kn}=0.0194,~N_{AB}=4$; (b) $\text{Kn}=0.1,~N_{AB}=6$. The analytical solution is included as the reference result for comparison.}
\label{uScominPois}
\end{figure}
In Fig. \ref{uScominPois}, the results predicted by the five cases of $l$ are shown and compared with the analytical solutions. Clearly, the results with the parameter $l$ conforming to Eq. \eqref{Eqexlva} exhibit the best agreement with the analytical solutions. However, clear deviations from the analytical solutions are observed for the other four cases of $l$. This is in expectation because the four choices of $l$ cannot generate uniform relaxation times $\tau_q$ to realize the slip boundary condition. These observations demonstrate the superiority of the present boundary scheme, as mentioned previously, over previous curved ones in capturing micro flows with curved walls.

\subsection{Couette flow between two concentric cylinders}
The proposed boundary scheme is further applied to a microgaseous flow with more complex geometries, i.e., the microcylindrical Couette flow between two cylinders, to validate the theoretical analysis. This classical problem has been studied by many researchers as a benchmark case in fluid dynamics \cite{Guo11,LiuZ19,Tao15,Lockerb04,Yuhong05}.
\begin{figure}
\centering
\includegraphics[height=0.45\textwidth,width=0.45\textwidth]{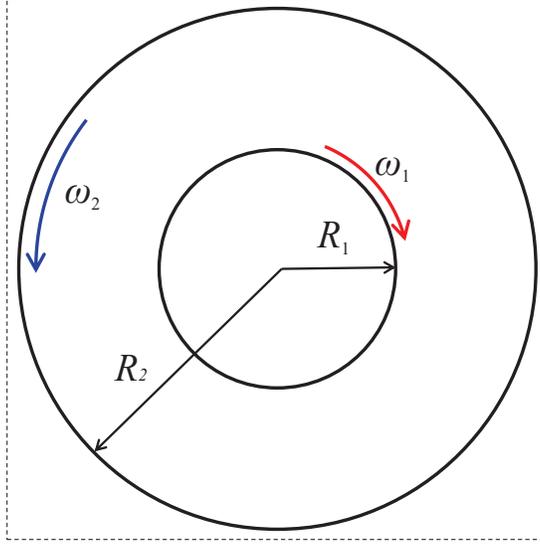}
\caption{Schematic illustration of the microcylindrical Couette flow.}
\label{ScheCouette}
\end{figure}
As shown in Fig. \ref{ScheCouette}, the two cylinders of the problem are concentric with respective radii of $R_1$ and $R_2$ ($R_1<R_2$), and rotate at angular velocities $\omega_1$ and $\omega_2$, respectively. As the flow between the cylinders reaches at steady state, they can be described by the following reduced Navier-Stokes equations in cylinder polar coordinates $(r, \theta)$
\begin{equation}\label{EqMicMC}
  \frac{\text{d}^2u_\theta}{\text{d} r^2}+\frac{\text{d}}{\text{d} r}\left(\frac{u_\theta}{r}\right)=0,
\end{equation}
where $u_\theta$ is the tangential velocity and $r$ is the radial distance. Based on the Maxwell's diffusive boundary condition for  gas-wall interactions, the slip boundary condition at the inner and outer cylinder walls can be expressed as
\begin{equation}\label{EQMCboun}
u_\theta\big|_{r=R_1}=\omega_1R_1+\frac{2-\sigma_1}{\sigma_1}\lambda\left(\frac{\text{d}u_\theta}{\text{d}r}-\frac{u_\theta}{r}\right)\bigg|_{r=R_1}, \quad u_\theta\big|_{r=R_2}=\omega_2R_2-\frac{2-\sigma_2}{\sigma_2}\lambda\left(\frac{\text{d}u_\theta}{\text{d}r}-\frac{u_\theta}{r}\right)\bigg|_{r=R_2},
\end{equation}
where $\sigma_1$ and $\sigma_2$ are the accommodation coefficients of the inner and outer cylinder surface, respectively. Under the boundary condition \eqref{EQMCboun}, the solution of Eq. \eqref{EqMicMC} for the velocity profile can be analytical solved as \cite{Yuhong05}
\begin{equation}\label{EqAnaMicro}
  u_\theta=\frac{\mathcal{M}\omega_1-\mathcal{N}\omega_2}{\mathcal{M}-\mathcal{N}}r
  +\frac{\omega_1-\omega_2}{\mathcal{N}-\mathcal{M}}\frac{1}{r},
\end{equation}
where
\begin{equation}\label{EqAnaPraMicro}
  \mathcal{M}=\frac{1}{R_2^2}\left(1-\frac{2-\sigma_2}{\sigma_2}\frac{2\lambda}{R_2}\right),\quad
  \mathcal{N}=\frac{1}{R_1^2}\left(1+\frac{2-\sigma_1}{\sigma_1}\frac{2\lambda}{R_1}\right).
\end{equation}

\begin{figure}
\begin{tabular}{cc}
\includegraphics[width=0.5\textwidth,height=0.25\textheight]{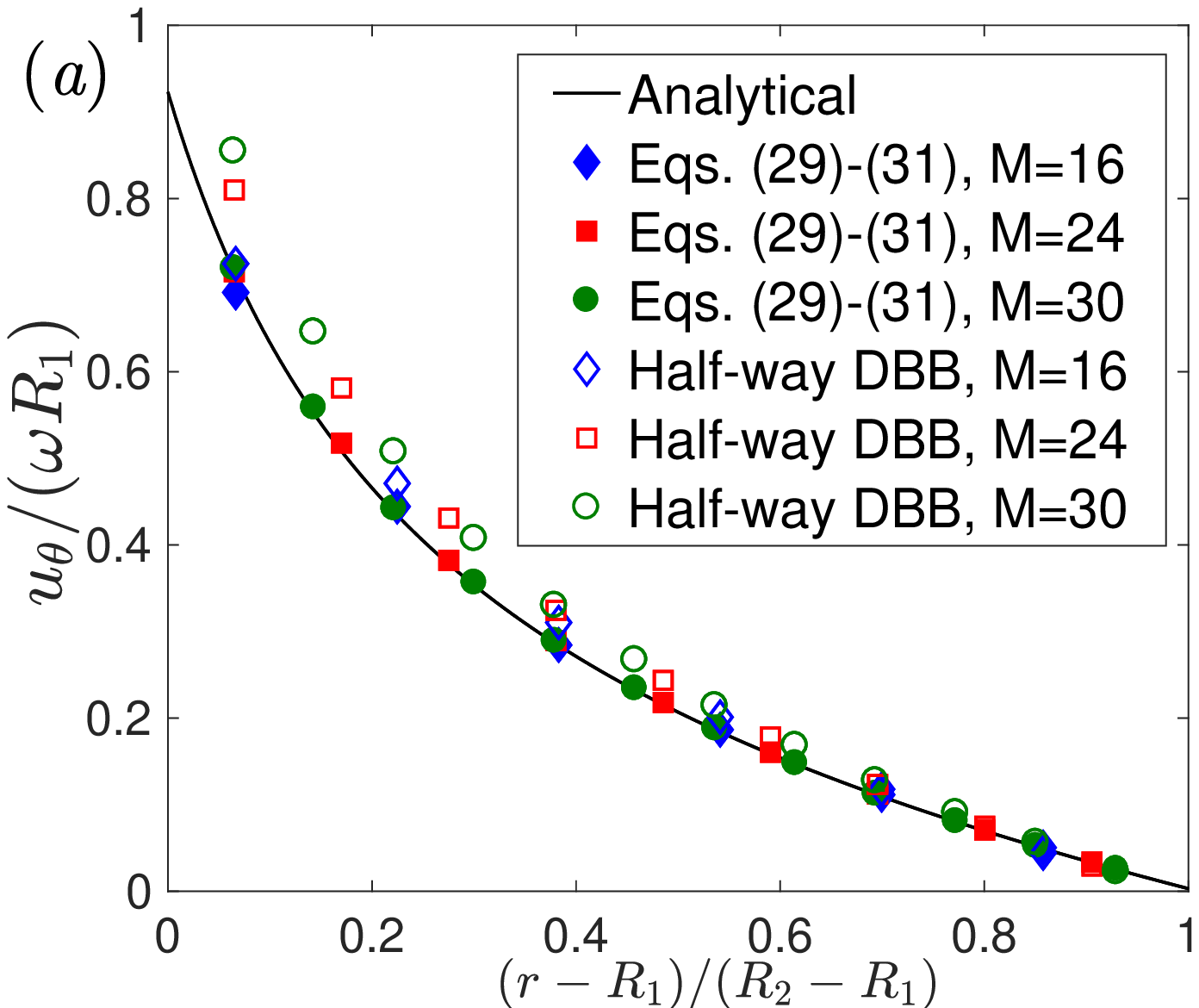}
\includegraphics[width=0.5\textwidth,height=0.25\textheight]{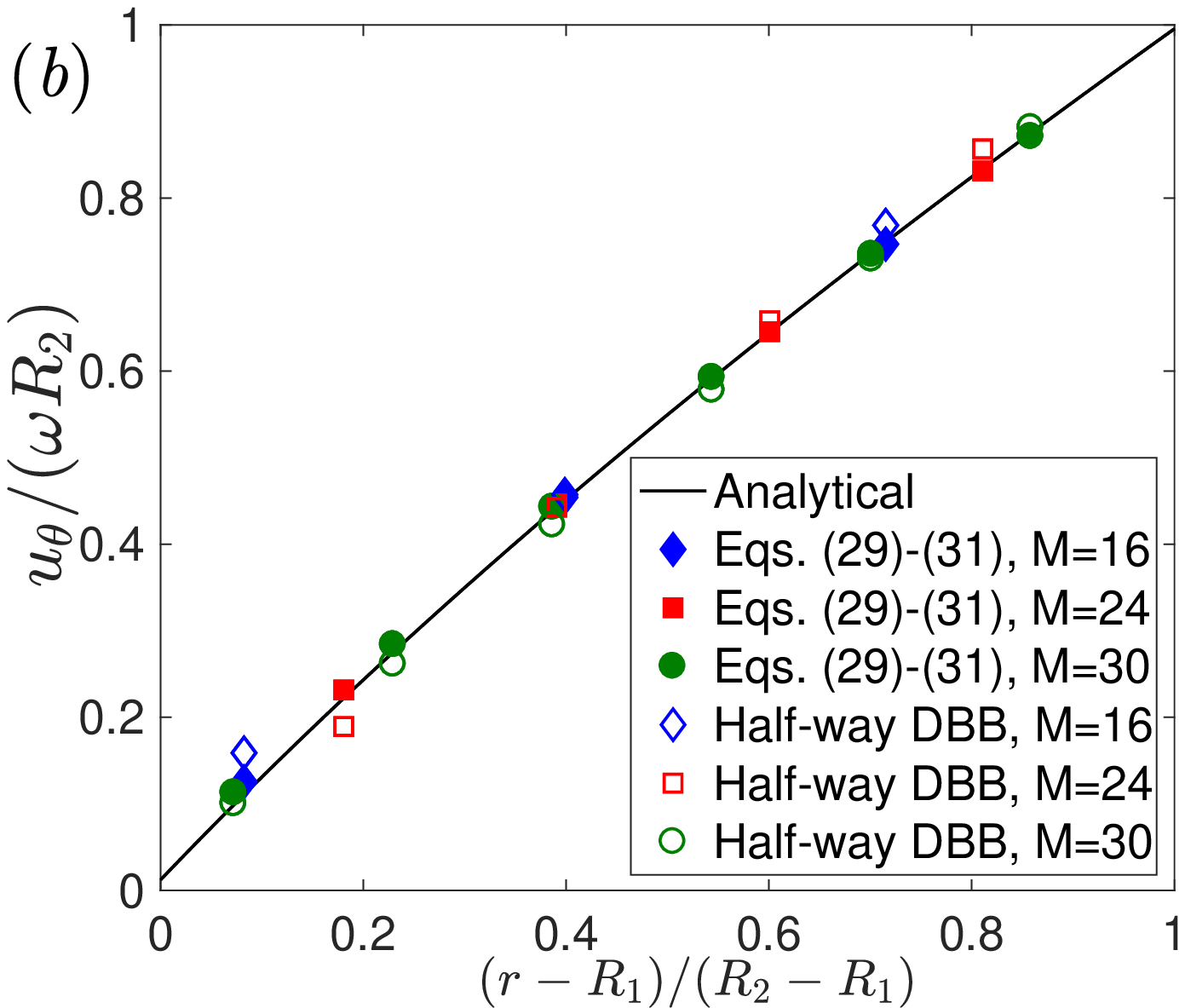}\\
\end{tabular}
\caption{Velocity profiles of the microcylindrical Couette flow for (a) $\beta=0.2,~\omega_1=0.001,~\omega_2=0$; (b) $\beta=0.6,~\omega_1=0,~\omega_2=0.001$. Filled shapes denote the results predicted by the boundary scheme \eqref{EqNewBDSC} with Eqs. \eqref{Eqexlva}-\eqref{NSliCoertstq} ($\mathcal{E}=-0.68$). Empty shapes are the results obtained by the halfway DBB scheme with Eqs. \eqref{SliCoeffHar} and \eqref{SliCoeffHata}.}
\label{uScominCoue}
\end{figure}

In the simulations, the two cylinder surfaces are both assumed to be fully diffusive, i.e., $\sigma_1=\sigma_2=\sigma=1$. To mimic the slip boundary condition \eqref{EQMCboun} at the two cylinders' surfaces, we set $\mathcal{L}_1=(2-\sigma)/2, ~\mathcal{L}_2=0$ in the boundary condition \eqref{EqNewBDSC} implemented with Eqs. \eqref{Eqexlva}-\eqref{NSliCoertstq}. The Knudsen number for the flow is defined as $\text{Kn}=\lambda/(R_2-R_1)$. The radius ratio of the two cylinders, $\beta=R_1/R_2$, is obtained by changing $R_1$ under the fixed $R_2=1.0$. The two cylinder's center is placed at that of a square domain covered by $M$ grid cells. Considering the small grid sizes used, we set $\text{Kn}=0.01$ and $\omega_1=\omega_2=\omega=0.001$ in the simulations. As done before, the value of $l$ is taken as zero in case $l<-0.4$ computed from Eq. \eqref{Eqexlva}, in which the constant $\mathcal{E}$ is assigned in the range of Eq. \eqref{EqECond}. In Fig. \ref{uScominCoue}, the tangential velocities with different lattice sizes are shown for two cases of rotating cylinders.
Clearly, the velocity profiles predicted by the boundary scheme \eqref{EqNewBDSC} with Eqs. \eqref{Eqexlva}-\eqref{NSliCoertstq} are in good agreement with the analytical solutions. Whereas for the halfway DBB scheme, Eqs. \eqref{SliCoeffHar} and \eqref{SliCoeffHata} cannot yield grid-independent results consistent with the analytical solutions. This is ascribed to the fact that the midway approximation for the cylinder surfaces distort the actual curved geometries. Our theoretical analyses on curved boundary conditions are demonstrated again here.

\begin{figure}
\begin{tabular}{cc}
\includegraphics[width=0.5\textwidth,height=0.25\textheight]{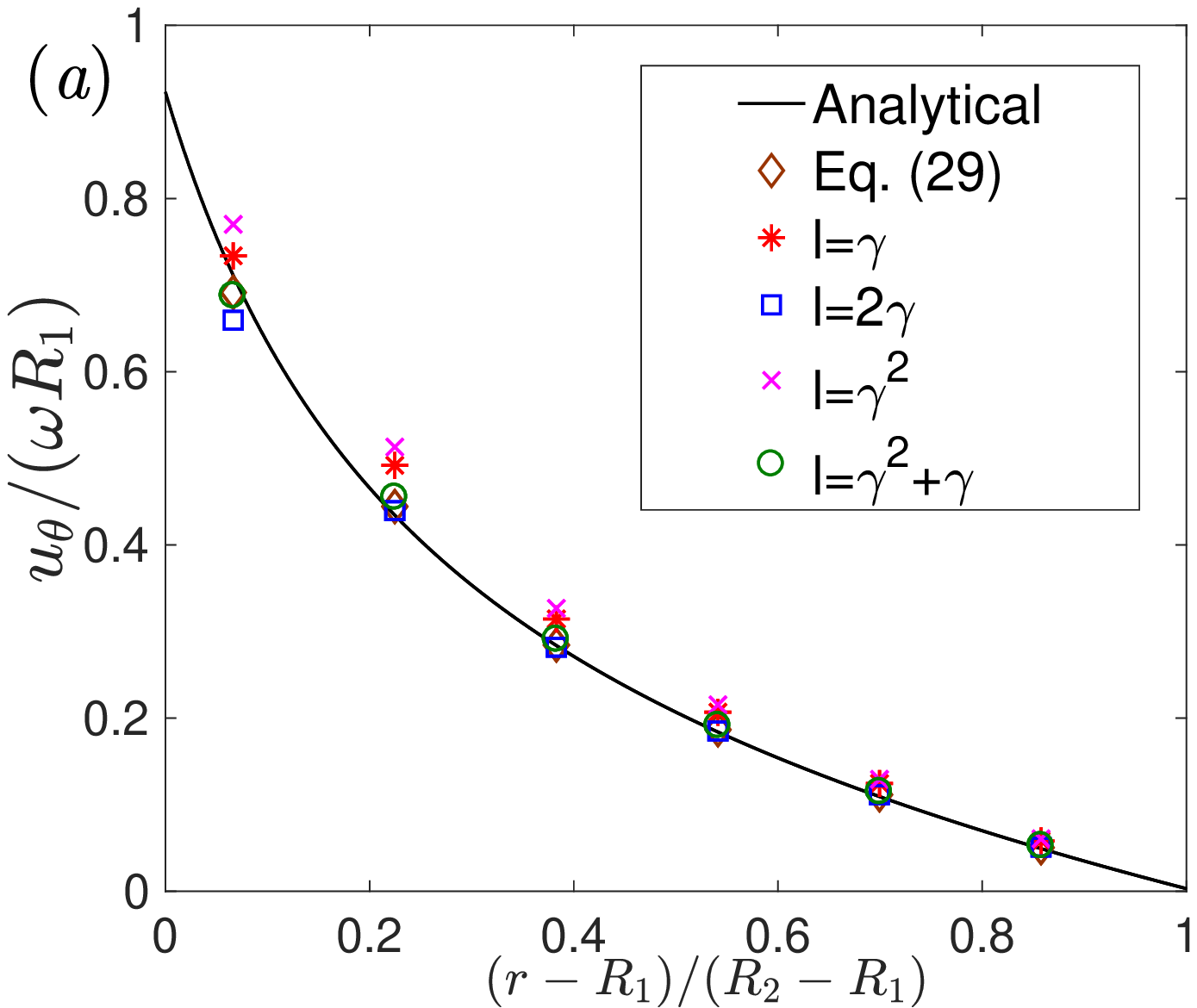}
\includegraphics[width=0.5\textwidth,height=0.25\textheight]{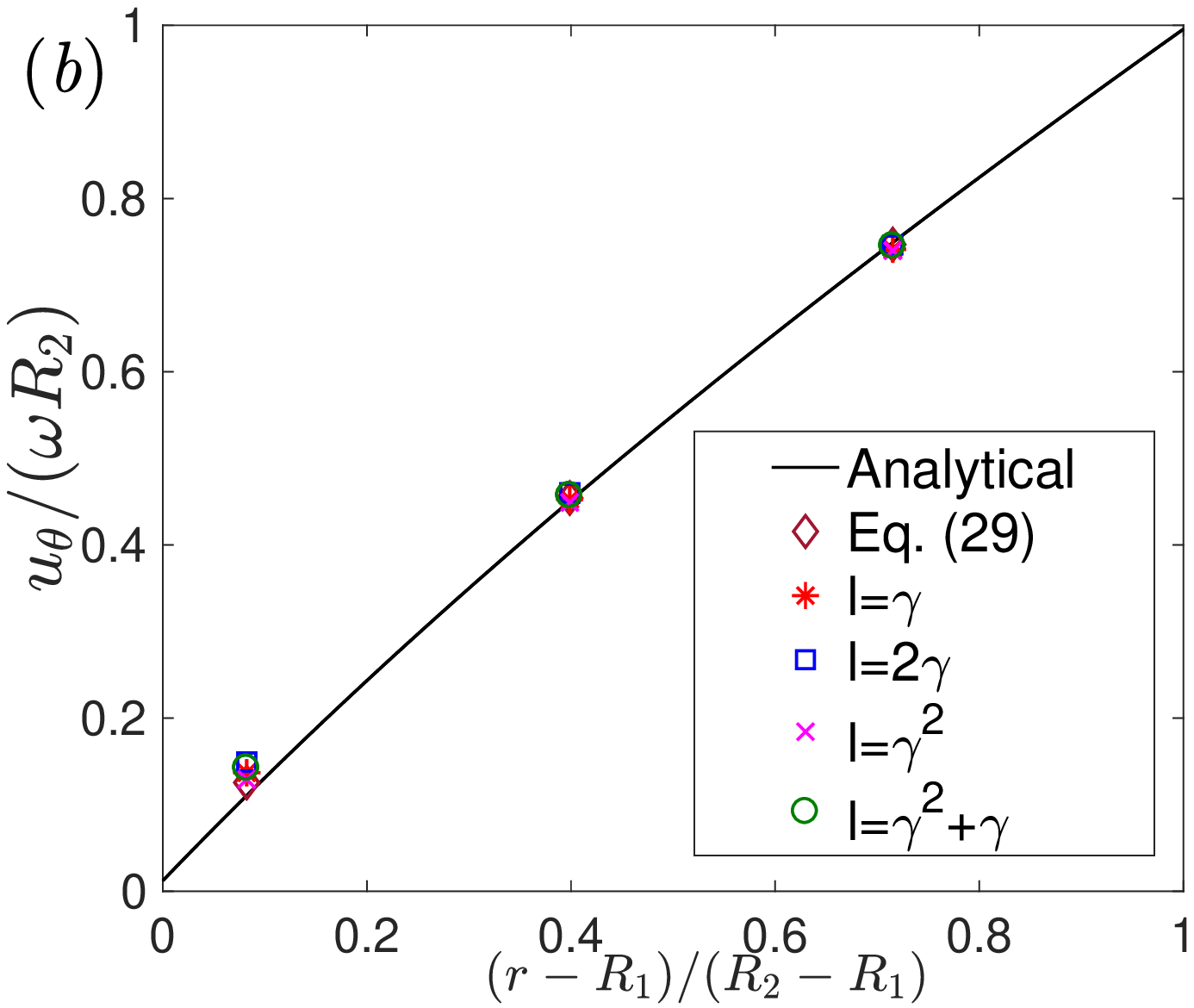}\\
\end{tabular}
\caption{Velocity profiles of the microcylindrical Couette flow for (a) $\beta=0.2,~\omega_1=0.001,~\omega_2=0$; (b) $\beta=0.6,~\omega_1=0,~\omega_2=0.001$ at $M=16$. }
\label{uScominMIcro}
\end{figure}

\begin{figure}
\begin{tabular}{cc}
\includegraphics[width=0.5\textwidth,height=0.25\textheight]{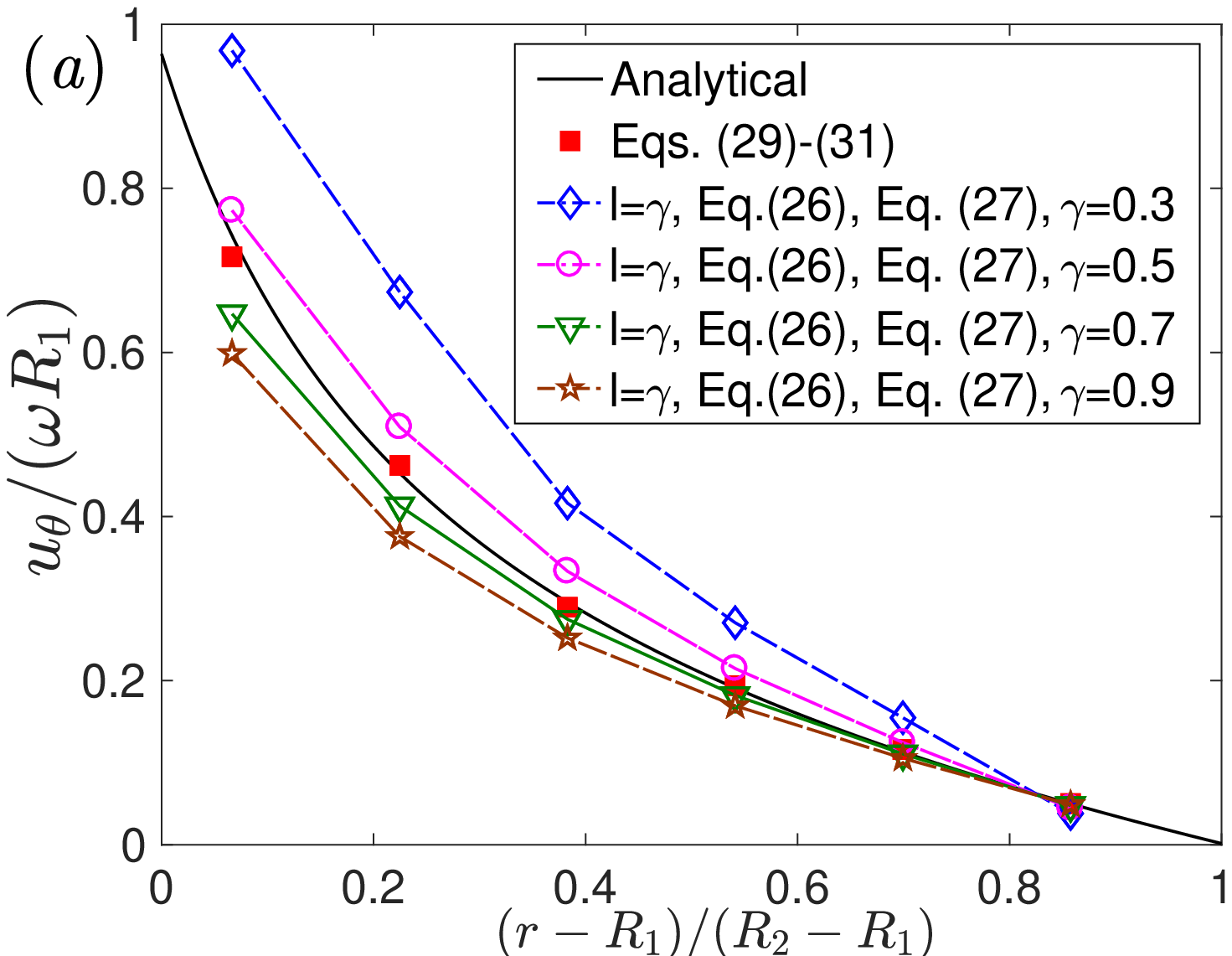}
\includegraphics[width=0.5\textwidth,height=0.25\textheight]{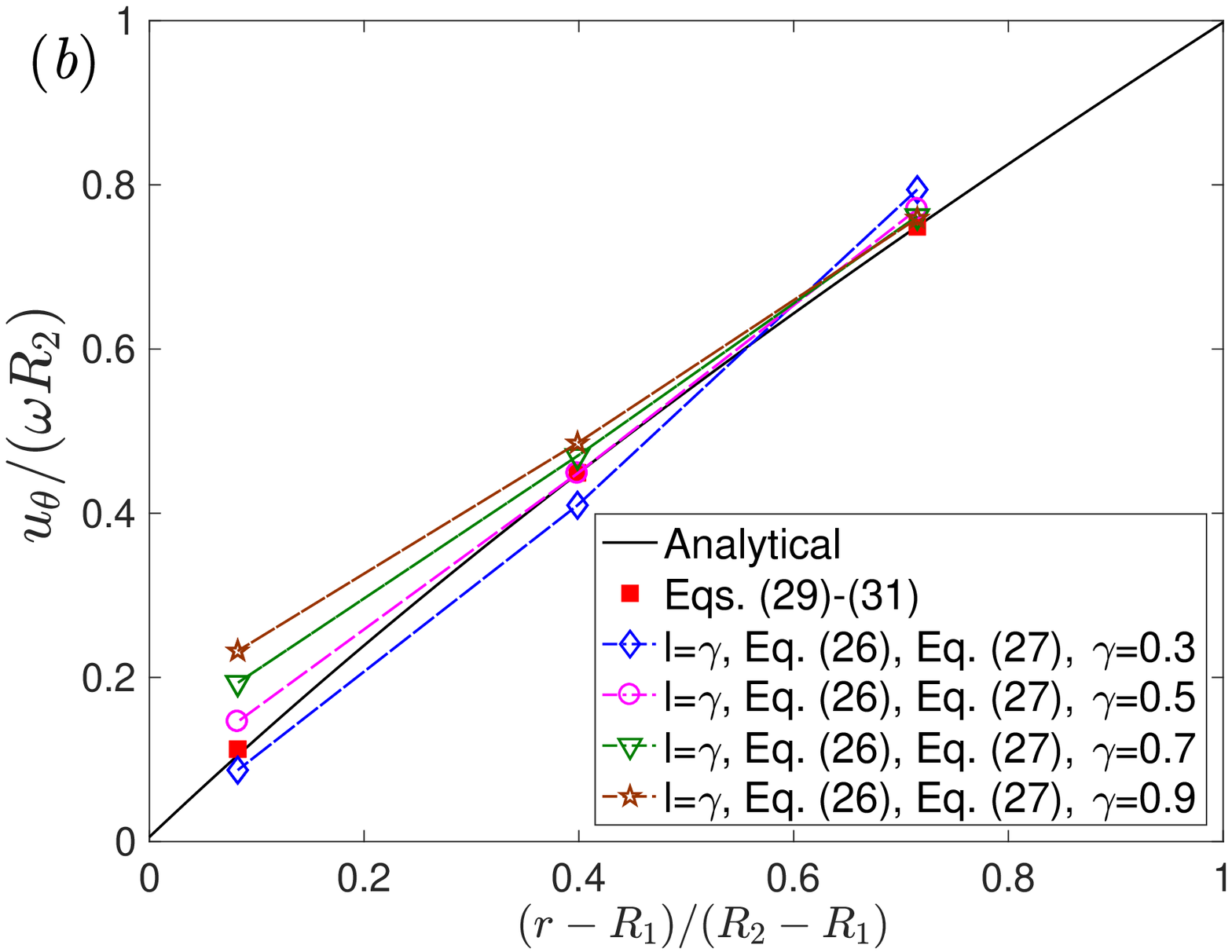}\\
\end{tabular}
\caption{Velocity profiles of the microcylindrical Couette flow at $\text{Kn}=0.0045$ and $M=16$ for (a) $\beta=0.2,~\omega_1=0.001,~\omega_2=0$;  (b) $\beta=0.6,~\omega_1=0,~\omega_2=0.001$. Filled square shapes denote the predicted results by the boundary scheme \eqref{EqNewBDSC} with Eqs. \eqref{Eqexlva}-\eqref{NSliCoertstq}. Empty shapes represent the results obtained by the boundary scheme \eqref{EqNewBDSC} with $l=\gamma$ in Eq. \eqref{SliCoeffrtsr} for $r$, and in Eq. \eqref{SliCoeffrtstq} to obtain the uniform $\tau_q$ by fixed values of $\gamma$.}
\label{uSMIcrogama}
\end{figure}

Figure \ref{uScominMIcro} displays the velocity profiles between the cylinders predicted under different cases of $l$. Four additional choices of $l$, i.e., $l=\gamma,~2\gamma,~\gamma^2,~\gamma^2+\gamma$ besides $l$ given by Eq. \eqref{Eqexlva} are considered in the simulations. In all these cases, the combination parameter $r$ is determined by Eq. \eqref{NSliCoertsr}, and $\tau_q$ by Eq. \eqref{NSliCoertstq} with $\mathcal{E}=-0.65$.
From the figure, it is observed that compared with the other choices of $l$, the optimal consistency with analytical solutions is obtained for the case that $l$ conforms to Eq. \eqref{Eqexlva}, even with the artificial value $l=0$ at several boundary points. This also indicates that the uniform relaxation time $\tau_q$ cannot be achieved for the other cases of $l$ to realize the slip boundary condition. In addition, the tangential velocity predicted with $l=\gamma$ in Eqs. \eqref{SliCoeffrtsr} and \eqref{SliCoeffrtstq} are investigated as $\gamma$ is artificially fixed to derive the uniform $\tau_q$. The simulated results are delineated in Fig. \ref{uSMIcrogama}, where $\text{Kn}=0.045$ is set to ensure $\tau_q>0.5$.
Again, apparent deviations from the analytical solutions are observed for the artificial approximation of $\gamma$. While with Eqs. \eqref{Eqexlva}-\eqref{NSliCoertstq}, the boundary scheme \eqref{EqNewBDSC} brings robust agreement results with the analytical solutions. These observations strengthen and confirm the capability of the present boundary scheme to realize the slip boundary condition at curved walls with uniform relaxation parameters.

\section{Conclusions} \label{Sec5}
In this work, a kinetic boundary condition has been developed for the LBM simulating microgaseous flows with curved geometries. This curved boundary scheme is a combination of the Maxwellian diffuse reflection scheme and a single-node boundary scheme for curved no-slip walls. In addition to the distance ratio $\gamma$, which is only involved in previous curved boundary treatments, an additional free parameter $l$ is incorporated in the present local boundary scheme. Based on the theoretical analysis within the framework of MRT model, it is shown that the free parameter $l$ as well as the distance ratio $\gamma$ and the relaxation times ($\tau_s$ and $\tau_q$) unitedly affect the slip velocity derived from the boundary scheme. Thanks to the free parameter $l$, an available strategy to determine the uniform $\tau_q$ together with the combination parameter $r$ is proposed to realize the slip boundary condition at solid walls. Furthermore, it is found that without the free parameter $l$, the accurate slip boundary condition cannot be realized with an invariable $\tau_q$ for the halfway DBB scheme and previous curved boundary schemes.

The proposed curved boundary condition is applied to some benchmark problems with planar and curved walls, including the aligned and inclined microchannel flows and the microcylindrical Couette flow. To avoid the instability at very small $\gamma$ in the simulations, the free parameter $l$ is artificially assigned with zero value at several boundary nodes. Good and robust predictions from the present derivations are obtained to match the analytical solutions satisfactorily even under a small lattice size. The numerical results also show that for microgaseous flows with nonplanar and curved walls, the halfway DBB scheme and previous curved boundary conditions which only contains $\gamma$ bring clear grid-dependent deviations from the analytical solutions.

On the basis of the present study, a noteworthy point is that to ensure uniform relaxation parameters in realizing a prescribed slip boundary condition at curved walls, adding free parameters to the boundary scheme would be more efficient than seeking more relaxation parameters in the LBE. It should be noted that this work on curved boundary conditions focused on microscale gas flows in the slip regime. However, the present analysis is also instructive for our extension work to the transition regime, where the effect of the Knudsen layer must be incorporated. In addition, we would like to point out that by replacing the Maxwellian diffusive part by the specular reflection scheme in Eq. \eqref{EqNewBDSC}, it can generate another hybrid but nonlocal boundary scheme for microscale flows.  Following the method presented in this work, a similar strategy for uniform relaxation parameters can be obtained to realize the prescribed slip boundary condition at curved walls. Furthermore, such idea of introducing a free parameter can be also applied to combining the Maxwellian diffusive boundary scheme and the specular reflection scheme. Finally, the present work can be extended to the three-dimensional case without much difficulty. These interesting topics will be left for our future work.

\vspace{5mm}
\begin{acknowledgments}
This work is supported by the National Natural Science Foundation of China (No. 51776068 and No. 51906044) and the Fundamental Research Funds for the Central Universities (No. 2018MS060). L. Wang would like to thank Profs. Wen-An Yong and Zhaoli Guo and Dr. Weifeng Zhao for their helpful discussions.
\end{acknowledgments}
%

\nocite{*}


\end{document}